\documentclass[11pt,oneside,a4paper]{article}

\usepackage[margin=2.5cm]{geometry}

\usepackage{setspace}
\usepackage{soul}
\usepackage{amsfonts,dsfont}
\usepackage{amsmath,amssymb}

\usepackage[square,numbers]{natbib}

\usepackage{setspace}

\usepackage{lipsum}
\newenvironment{localsize}[1]
{%
  \clearpage
  \let\orignewcommand\newcommand
  \let\newcommand\renewcommand
  \makeatletter
  \input{bk#1.clo}%
  \makeatother
  \let\newcommand\orignewcommand
}

\usepackage[titletoc,title]{appendix}

\usepackage[ruled]{algorithm2e}

\SetAlFnt{\small}
\SetAlCapFnt{\small}
\SetAlCapNameFnt{\small}
\SetAlCapHSkip{0pt}
\IncMargin{-\parindent}

\usepackage{graphicx}
\graphicspath{ {./} }
\DeclareGraphicsExtensions{.pdf}

\title{\textbf{Optimizing the Convergence Rate of the Continuous Time Quantum Consensus}}
\author{Saber Jafarizadeh  \\ {saber.jafarizadeh@sydney.edu.au} }
\date{}
\providecommand{\keywords}[1]{\textbf{\textit{Index terms---}} #1}

\usepackage{fancyhdr}

\begin{document}

\maketitle

\markboth{S. Jafarizadeh}{Continuous Time Quantum Consensus}

\bibliographystyle{ieeetran}

\begin{abstract}

Inspired by the recent developments in the fields of quantum distributed computing, quantum systems are analyzed as networks of quantum nodes to reduce the complexity of the analysis.
This gives rise to the distributed quantum consensus algorithms.
Focus of this paper is on optimizing the convergence rate of the continuous time quantum consensus algorithm over a quantum network with $N$ qudits.
It is shown that the optimal convergence rate is independent of the value of $d$ in qudits.
First by classifying the induced graphs as the Schreier graphs, they are categorized in terms of the partitions of integer $N$.
Then establishing the intertwining relation between one level dominant partitions in the Hasse Diagram of integer $N$, it is proved that the spectrum of the induced graph corresponding to the dominant partition is included in that of the less dominant partition.
Based on this result, the proof of the Aldous' conjecture is extended to all possible induced graphs  and the original optimization problem is reduced to optimizing spectral gap of the smallest induced graph.
By providing the analytical solution to semidefinite programming formulation of the obtained problem, closed-form expressions for the optimal results are provided for a wide range of topologies.

\end{abstract}


\section{Introduction}

Coordinated control and consensus are the essential factors in coupled dynamical systems that are modeled as networks of autonomous agents.
Examples of such systems in nature are flocks of birds, school of fish, neurons within the nervous system \cite{KocarevBookComplexConsensus2013}.
Other manmade examples include power grids and social networks \cite{WeiBookVehicleConsensus2008}.
Within the fields of distributed control and optimization on networks, there have been remarkable efforts in modeling and controlling cooperative collective behaviors in networks of autonomous agents.
One of the basic behaviors of autonomous agents is reaching consensus, where agents in a network achieve a common state using only local communication with each other \cite{Olfati2004,Jadbabaie2003,Tsitsiklis1986}.

Due to recent progress in the fields of quantum information science and quantum distributed computing \cite{Broadbent2008,Buhrman2003,Denchev:2008}, quantum systems are analyzed as networks of Quantum nodes.
This is because building and maintaining a centralized and rather big quantum computer with many qubits can be excessively difficult and expensive.
As an alternative solution, such machine can be replaced by a network of smaller quantum computers to carry out computation on a scale that would yield practical benefits.

In \cite{PetersenRef15,MazzarellaCDC2013,MazzarellaArXiv}  authors extended the consensus problem to the quantum domain by reinterpreting it as a symmetrization problem.
They have addressed this problem by a switching-type dynamics based on convex combinations of actions of a finite group.
Furthermore, they derive the general conditions for convergence and show that convergence is guaranteed provided that some mild assumptions are held.
Using the results on convergence, they prove that it ensures asymptotic convergence as well.

Authors in \cite{Petersen2015IEEETranAutControl,Petersen2015ACCPartI,Petersen2015ACCPartII} propose a new approach based on the induced graphs of the quantum interaction graph for relating the quantum consensus over the $N$-qubit network to the classical consensus dynamics.
They have shown how to carry out convergence speed optimization of the equivalent classical consensus via convex programming.
Furthermore, they establish necessary and sufficient conditions for exponential and asymptotic quantum consensus, respectively, for switching quantum interaction graphs.

In the present paper, we optimize the convergence rate of the quantum consensus over a quantum network with $N$ qudits.
The main motivation for extending this problem from qubits to qudits is to show that the optimal convergence rate is independent of the value of $d$ in qudits.
By expanding the density matrix in terms of the generalized Gell-Mann matrices, we have shown that the induced graphs are the Schreier graphs where in the special case of interchange process it reduces to the Cayley graph.
Then using the Young Tabloids, we categorize the induced graphs obtained from all possible partitions of the integer $N$.
By establishing the intertwining relation between one level dominant partitions in the Hasse Diagram of integer $N$, we have shown that the eigenvalues of the induced graph corresponding to the dominant partition is included in the eigenvalues of the less dominant partition.
Based on this result, we extend the proof of the Aldous' conjecture \cite{ProofAldous} to all possible induced graphs
and we show that the problem of optimizing the convergence rate of the quantum consensus reduces to optimizing the second smallest eigenvalue of the Laplacian of the induced graph corresponding to partition $(N-1,1)$.
Based on the extension of Aldous' conjecture, it is shown that if one of the induced graphs serves as the underlying graph for another quantum network then the spectral gap and thus the convergence rate for induced graphs obtained from the new network is same as those of the old one.
In this way, a numerous number of weighted Laplacian matrices with the same spectral gap can be obtained.
In the final stage, we have analytically solved the semidefinite programming formulation of the reduced optimization problem for a wide range of topologies and provided closed-form expressions for the optimal convergence rate and the optimal weights.

The rest of the paper is organized as follows.
Section \ref{Preliminaries} presents some preliminaries including relevant concepts in graph theory, Young tabloids, Cayley and Schreier Coset graphs.
The continuous time consensus algorithm and the semidefinite programming formulation of its optimization are presented in Section \ref{sec:CTCPriliminaries}.
Section \ref{sec:ContinuousTimeQuantumConsensus} describes optimization of the continuous time quantum consensus problem and how it can be transformed into optimization of a classical continuous time consensus problem.
In Section \ref{sec:Lambda2SDP} analytical optimization of the continuous time consensus problem and closed-form expressions for the optimal results for a range of topologies have been presented.

\section{Preliminaries}
\label{Preliminaries}

In this section, we present the fundamental concepts from graph theory, Young tabloids, Hasse diagrams, Cayley and Schreier coset graphs.

\subsection{Graph Theory}
\label{sec:GraphTheory}

A graph is defined as $\mathcal{G}= \{ \mathcal{V}, \mathcal{E} \}$ with $\mathcal{V} = \{1, \ldots, N\}$ as the set of vertices and $\mathcal{E}$ as the set of edges.
Each edge $\{i,j\} \in \mathcal{E}$ is an unordered pair of distinct vertices.
If no direction is assigned to the edges, then the graph is called an undirected graph.
Throughout this paper, we consider undirected simple graphs with no self-loops and at most one edge between any two different vertices.
The set of all neighbors of a vertex $i$ is defined as $\mathcal{N}_{i}  \triangleq  \{  j \in \mathcal{V} : \{i,j\} \in \mathcal{E}  \}$.
A weighted graph is a graph where a weight is associated with every edge according to proper map $W: \mathcal{E} \rightarrow \mathbb{R}$, such that if $\{i,j\} \in \mathcal{E}$, then $W(\{i,j\})= \boldsymbol{w}_{ij}$; otherwise $W(\{i,j\}) = 0$.
The edge structure of the weighted graph $\mathcal{G}$ is described through its adjacency matrix $(A_{\mathcal{G}})$.
The adjacency matrix $A_{\mathcal{G}}$ is a $N \times N$ matrix with $\{i,j\}$-th entry $(A_{\mathcal{G}}(i,j))$ defined as below
\begin{equation}
    \nonumber
    \begin{gathered}
        \nonumber  A_\mathcal{G}(i,j) =
	       \begin{cases}
                \boldsymbol{w}_{ij} \quad \text{if} \quad \{i,j\} \in \mathcal{E} \\
                0 \quad \text{Otherwise}
            \end{cases}
     \end{gathered}
\end{equation}
i.e., the $(i,j)-{th}$ entry of $A_{\mathcal{G}}$ is $1$ if vertex $j$ is a neighbor of vertex $i$.
If the graph $\mathcal{G}$ has no self-loops $A_{\mathcal{G}}(i,i) = 0$, i.e., the diagonal elements of the adjacency matrix are all equal to zero.
For undirected graphs the adjacency matrix is symmetric, i.e., $A_{\mathcal{G}}$ is symmetric.
The degree of a vertex $i$ is the sum of the weights on the edges connected to vertex $i$, i.e.
\begin{equation}
    \nonumber
    \begin{gathered}
        d_{i} = \sum_{j=1}^{N} {\boldsymbol{w}_{ij}}.
     \end{gathered}
\end{equation}
The degree matrix $D_{\mathcal{G}}$ of $\mathcal{G}$ is the $N \times N$ diagonal matrix where its $i$-th diagonal element is equal to the degree of vertex $i$ and all non-diagonal elements are equal to zero.
A graph is called connected if there is a path between any two vertices in the graph.
A graph is called a regular graph if all the vertices have the same number of neighbors.
The Laplacian matrix of graph $\mathcal{G}$ is defined as below,
\begin{equation}
    \nonumber
    \begin{gathered}
        \nonumber  L_{\mathcal{G}}(i,j) =
	       \begin{cases}
                D_{\mathcal{G}}(i,i) \quad \text{if} \quad i = j \\
                -A_{\mathcal{G}}(i,j) \quad \text{if} \quad i \neq j
            \end{cases}
     \end{gathered}
\end{equation}
This definition of the Laplacian matrix can be expressed in matrix form as $L_{\mathcal{G}} = D_{\mathcal{G}} - A_{\mathcal{G}}$, where $D_{\mathcal{G}}$ and $A_{\mathcal{G}}$ are the degree and the adjacency matrices of the graph $\mathcal{G}$.
The Laplacian matrix of an undirected graph is a symmetric matrix.
The eigenvalues of the Laplacian matrix $(L_{\mathcal{G}})$ are all nonnegative.
Defining $\bf{1}$ and $\bf{0}$ as vectors of length $N$ with all elements equal to one and zero, respectively, hence for the Laplacian matrix we have $L_{\mathcal{G}} \times \bf{1} = \bf{0}$.
In undirected graphs, the associated Laplacian is a positive semidefinite matrix and its eigenvalues can be arranged in non-decreasing order as follows,
\begin{equation}
    \nonumber
    \begin{gathered}
        0 = \lambda_{1} (L_{\mathcal{G}}) \leq \lambda_{2} (L_{\mathcal{G}}) \leq \cdots \leq \lambda_{N} (L_{\mathcal{G}})
     \end{gathered}
\end{equation}
The second smallest eigenvalue $\lambda_{2} (L_{\mathcal{G}})$ is known as the algebraic connectivity and reflects the degree of connectivity of the graph \cite{Fiedler1973}.
First introduced in \cite{Fiedler1973}, this eigenvalue is named algebraic connectivity due to its importance in connectivity properties of the graph.
Since then the algebraic connectivity has found applications in the analysis of numerous problems including combinatorial optimization problems such as the maximum cut problem, certain flowing process and the traveling salesman problem \cite{BrazilReview2007}.
The algebraic connectivity can be used to define the spectral gap.
The spectral gap gives insight into important properties of the graph such as the mixing time of random walks \cite{Hoory2006}.
In some cases, the term spectral gap is directly used to refer to $\lambda_{2}(L_{\mathcal{G}})$.
A necessary and sufficient condition for the algebraic connectivity to be nonzero is that the graph $\mathcal{G}$ is connected \cite{Chung1997}.
If the algebraic connectivity of the graph $\mathcal{G}$ is nonzero then $L_{\mathcal{G}}$ is an irreducible matrix i.e. it is not similar to a block upper triangular matrix with two blocks via a permutation \cite{Horn2006}.
The largest eigenvalue $\lambda_{N} (L_{\mathcal{G}})$ of the Laplacian matrix is known as the Laplacian spectral radius of $\mathcal{G}$.

\subsection{Symmetric Group, Young Tabloids, Young subgroup \& Hasse Diagrams}
\label{sec:YoungHasse}
The set of all bijections $\Pi: \{1, \cdots , N\}\rightarrow \{1, \cdots , N\}$ with composition of maps forms a finite group of order $N!$, called symmetric or permutation group  denoted by $S_N$.
A standard notation for the permutation that sends $i\rightarrow \Pi(i)$ is:
\begin{equation}\label{permutation}
  \binom{1 \hspace{17pt}  2  \hspace{19pt}   3 \hspace{8pt} \cdots \hspace{8pt} N} { \Pi(1) \hspace{4pt} \Pi(2) \hspace{4pt} \Pi(3)\cdots\Pi(N)}
\end{equation}
A $r$-cycle is a permutation of the form 
$\Pi(l_{i}) = l_{i+1}$ for $i=1,\ldots, r-1$ and $\Pi(l_{r}) = l_{1}$ 
where $l_1,\cdots, l_r \in \{1,\cdots,N\}$ are distinct from each other and $\Pi(i) = i$ if $i$ not among the $l_j$.
The standard notation for this cycle is $(l_1, l_2,l_3,\cdots, l_r)$.
A transposition is a cycle of length $2$, and an elementary transposition is a transposition of the form $(i, i + 1)$.
Every permutation $\Pi\in S_N$ can be written as a product of disjoint cycles and cycles can be written as a product of elementary transpositions.

\begin{figure}
  \centering
     \includegraphics[width=130mm]{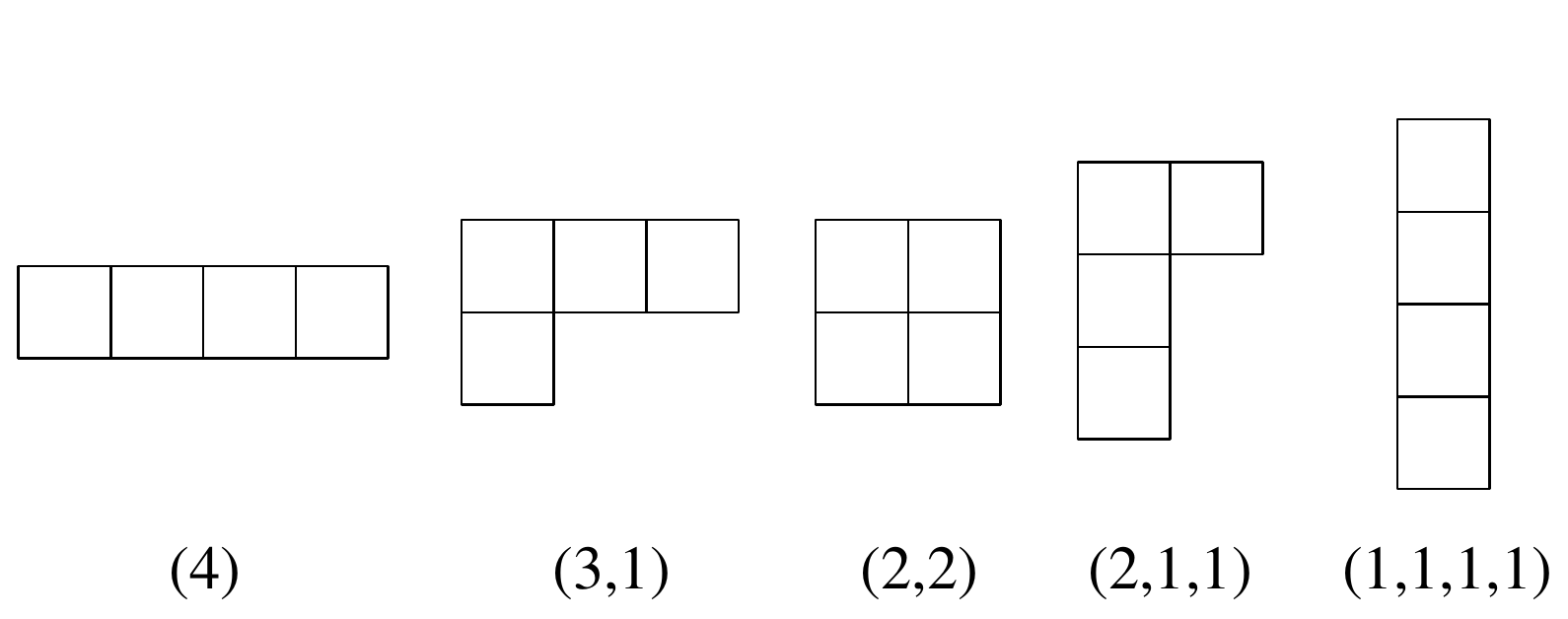}
  \caption{Young diagram for all possible partitions of $4$.}
  \label{fig:YoungDiagramExamplefor4}
\end{figure}

A Positive integer $N$ can be partitioned into a group of positive integers $n=(n_1, n_2, \ldots, n_K)$ where their summation is equal to $N$ and they are sorted in non-increasing order, i.e. $n_1 \geq n_2 \geq \cdots \geq n_K$.
$n$ is referred to as a partition of $N$ and it is denoted by $n \vdash N$.
A Young diagram is a finite set of boxes arranged in left-justified rows with non-increasing lengths.
In the corresponding Young diagram of the partition $n=(n_{1}, n_{2}, \ldots, n_{K})$, there are $n_{i}$ boxes in the $i$-th row of the diagram.
As an example, the possible partitions for number $4$ are $(4), (3,1), (2,2), (2,1,1), (1,1,1,1)$ and their corresponding Young diagrams are depicted in figure \ref{fig:YoungDiagramExamplefor4}.
It is obvious that there is a one-to-one correspondence between partitions and the Young diagrams.
For a given partition $n \vdash N$, a Young tableau of $n$-shape is obtained by filling in the boxes of the corresponding Young diagram of partition $n$ with integers from $1$ to $N$.
For a given partition if the integers in rows and columns are ordered in increasing order, then the Young tableau is referred to as standard Young tableau.
In total for a given Young diagram, there are $N!$ Young tableaux.
The definitions above for Young tableau and diagram are adapted from \cite{Segan2001}.

A Young tabloid is an equivalence class of Young tableau under the relation that two tableau are equivalent if each row contains the same elements.
The notation used for the Young tabloid is similar to the Young tableau but without vertical bars separating the entries within each row.
For a given partition $n$, the number of Young Tabloids is equal to ${\nu=N!/(n_{1}! \cdot n_{2}! \cdots  n_{K}!)}$.

Young tabloids of a given partition $n$ can be characterized in simple way by so called Yamanouchi symbols. 
For any  Young tabloid of $n$-shape we define a Yamanouchi symbol as a row of $N$ numbers $(r_1, r_{2},\cdots, r_{N-1}, r_{N})$ where
$r_i$ is the row in which the $i$-th number appears in the Young tabloid.
Based on Yamanouchi symbol we use the notation $t_{n}(r_{1}, r_{2},\cdots, r_{N-1}, r_{N})$ to uniquely represent a Young tabloid of partition $n$.

\begin{figure}
  \centering
     \includegraphics[width=30mm]{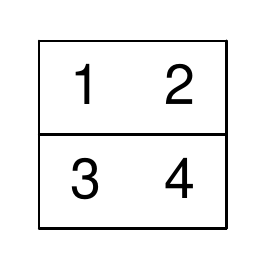}
  \caption{Young tabloid for partition $(2,2)$.}
  \label{fig:YoungTabloidExamplefor4}
\end{figure}

\begin{figure}
  \centering
     \includegraphics[width=130mm]{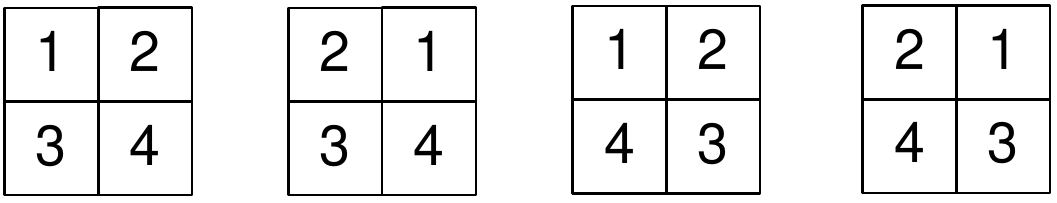}
  \caption{Four possible tableaux for partition $(2,2)$.}
  \label{fig:YoungTableauxExamplefor4}
\end{figure}

As an example, the Young tabloid depicted in figure \ref{fig:YoungTabloidExamplefor4} for partition $(2,2)$ represents the equivalence class containing the four tableaux presented in figure \ref{fig:YoungTableauxExamplefor4}.

\begin{figure}
  \centering
     \includegraphics[width=60mm]{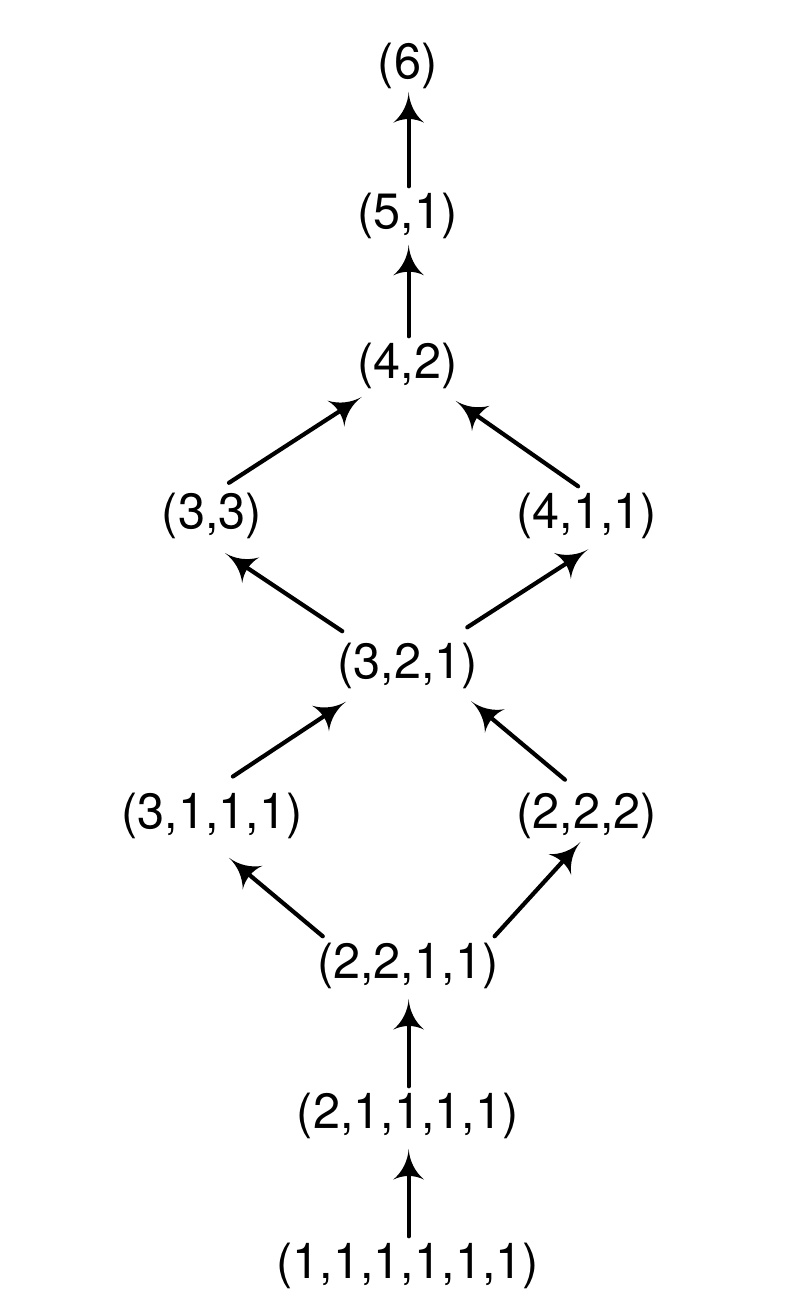}
  \caption{The Hasse diagram for all possible partitions of $T=6$.}
  \label{fig:HasseDiagram}
\end{figure}

Let $n=[n_1, n_2, \ldots]$ and $n^{'}=[ n_1^{'}, n_2^{'}, \ldots]$ be two given partitions of $N$ (i.e. $n \vdash N$ and $n^{'} \vdash N$) then $n$ dominates $n^{'}$ if for all $i \geq 1$, the sum of $i$ greatest parts of $n$ is greater than or equal to the sum of $i$ greatest parts of $n^{'}$. In other words,
\begin{equation}
    \nonumber
    \begin{aligned}
        n \unrhd n^{'} \quad \text{if and only if} \quad \sum_{j=1}^{i} {n_{j}} \geq \sum_{j=1}^{i} {n_{j}^{'}}  \quad \text{for all} \quad i \geq 1.
    \end{aligned}
\end{equation}
Note that in above definition of partition dominance, partitions $n$ and $n^{'}$ are extended by additional zero parts at the end as necessary.
The dominance relation between two sequences of numbers is also known as majorization  \cite{MajorizationBookRefMarshall}.
In terms of the Young diagrams, the number of squares in the first $i$ rows of the Young diagram of partition $n$ is greater or equal to that of partition $n^{'}$.
The diagram for dominance relations between partitions of a given number is known as the Hasse diagram, and it is used to represent partially ordered sets.
As an example for $N=6$, partition $(3,3)$ dominates partition $(2,2,1,1)$ but partitions (3, 3) and (4, 1, 1) are incomparable, since neither dominates the other.
The Hasse diagram for all possible partitions of $N=6$ is depicted in figure \ref{fig:HasseDiagram}.

For $n \vdash N$ , the $M^n$ is the vector space over real numbers $\mathbb{R}$ whose basis consists of a set of tabloids of $n$-shape given by
\begin{equation}
    \begin{gathered}
        \nonumber
        M^n=R\{\{t_n(r_{1}(1), r_{2}(1), \cdots, r_{N}(1))\},\cdots,\{t_n(r_{1}(\nu), r_{2}(\nu), \cdots, r_{N}(\nu))\}\}
    \end{gathered}
\end{equation}
where the set $\{\{t_n(r_{1}(1), r_{2}(1), \cdots, r_{N}(1))\},\cdots,\{t_n(r_{1}(\nu), r_{2}(\nu), \cdots, r_{N}(\nu))\}\}$ is a complete list of distinct tabloids of $n$-shape.
The symmetric group $S_{N}$ acts transitively over this set, i.e. by permutation the Yamanouchi symbols of a tabloid can be transformed to those of any other tabloid from the same set.
Thus $M^n$ is a representation of $S_N$ called the permutation module corresponding to $n$.

For $n=(n_1,n_2,\cdots,n_K)\vdash N$, the Young subgroup of $S_{N}$ corresponding to $n$ is defined as $S_n \overset{def}{=} S_{n_{1}}\times S_{n_{2}}\times\cdots \times S_{n_{K}}$, where $S_{n_{1}}$ permutes $1,2,\cdots,n_1$, $S_{n_{2}}$ permutes $n_1+1, n_1+2, \cdots ,n_1+n_2$ and so on.
The order of the Young subgroup of $n$-shape is $n_1!n_2!\cdots n_K!$.
Since $S_n$ is a subgroup of $S_N$, the number of left or right cosets of $S_n$  in $S_N$ is $N!/(n_1!\cdot n_2!\cdots n_K!)$ which is also   number of distinct tabloids of $n$-shape or $dim(M^n) $, hence there is a bijection between $\Pi_i S_n$ and the  $\{ \Pi_i t^n\}$, where $\{\Pi_i\}$ is a transversal for $S_n$ in $S_N$.

\subsection{Cayley Graph \& Schreier Coset Graph}

Let $\mathcal{H}$ be a group and let $\mathcal{S} \subseteq \mathcal{H}$.
The Cayley graph of $\mathcal{H}$ generated by $\mathcal{S}$ (referred to as the generator set $\mathcal{S}$), denoted by $Cay(\mathcal{H}, \mathcal{S})$, is the directed graph $\mathcal{G} = (\mathcal{V}, \mathcal{E})$ where $\mathcal{V} = \mathcal{H}$ and $\mathcal{E} = \{(x, xs) | x \in \mathcal{H}, s \in \mathcal{S}\}$.
If $\mathcal{S} = \mathcal{S}^{-1}$ (i.e., $\mathcal{S}$ is closed under inverse), then $Cay(\mathcal{H}, \mathcal{S})$ is an undirected graph.
If $\mathcal{H}$ acts transitively on a finite set $\Omega$, we may form a graph with vertex set $\mathcal{V} = \Omega$ and edge set $\mathcal{E} = \{ (\nu, \nu s) | \nu \in \Omega, s \in \mathcal{S} \}$. Similarly, if $\mathcal{Q}$ is a subgroup  in $\mathcal{H}$, we may form a graph whose vertices are the right cosets of $\mathcal{Q}$ , denoted $(\mathcal{H}:\mathcal{Q})$ and whose edges are of the form $\mathcal{E} = \{(\mathcal{Q}h, \mathcal{Q}hs) |\mathcal{Q}h\in (\mathcal{H}:\mathcal{Q}), s \in \mathcal{S}\}$.
These two graphs are the same when $\Omega$ is the coset space $(\mathcal{H}:\mathcal{Q})$, or when $\mathcal{Q}$ is the stabilizer of a point of $\Omega$ and  is called the Schreier coset graph $Sch(\mathcal{H}, \mathcal{S}, \mathcal{Q})$.

\section{Classical Continuous Time Consensus (CTC)}
\label{sec:CTCPriliminaries}

Consider a group of $N$ agents with an underlying connected graph $\mathcal{G} = (\mathcal{V}, \mathcal{E})$.
Each edge $\{i,j\}$ indicate bidirectional communication between agent $i$ and agent $j$, thus the resultant underlying graph $\mathcal{G}$ is an undirected graph.

Let $x_{i}$ be the state of agent $i$.
In the continuous time consensus (CTC) algorithm, each agent's dynamics evolves according to the following state dynamics equation,
\begin{equation}
    \label{eq:ConsensusStateDynamicsEquation}
    \begin{aligned}
        \dot{x}_{i}(t) = \sum_{j=1}^{N} { \boldsymbol{w}_{ij} (x_{i}(t) - x_{j}(t)) }, \quad \text{for} \quad i=1, \ldots, N,
    \end{aligned}
\end{equation}
Defining the vector $x = [x_{1}, \cdots, x_{N}]^{T}$ as the vector of states, we can rewrite the above state evolution formula in compact notation as below,
\begin{equation}
    \label{eq:CTCStateUpdate}
    \begin{aligned}
        \dot{x}(t) = -L_{\mathcal{G}} \times x(t), \quad t \in \mathbb{R} \geq 0,
    \end{aligned}
\end{equation}
where $L_{\mathcal{G}}$ is the graph Laplacian matrix for a weighted graph as defined in section \ref{sec:GraphTheory}.
It is well-known that $ \lim_{t \rightarrow \infty}{e^{-L_{\mathcal{G}}t}} \rightarrow \bf{1} \times \bf{1}^{T} / N $ where $\bf{1}$ is the left eigenvector of $L_{\mathcal{G}}$ corresponding to eigenvalue $0$.
Thus according to the state dynamics equation (\ref{eq:ConsensusStateDynamicsEquation}) it can be concluded that $\lim_{t \rightarrow \infty}{x(t)} \rightarrow  \frac{1}{N}\bf{1}\times \bf{1}^{T} \times x(0) = \frac{1}{N} \sum\nolimits_{j}{x_{j}(0)}\bf{1}$.
In other words, the final equilibrium state of the consensus algorithm is the average of agents' initial states if the underlying graph is connected.
The convergence rate of the algorithm to its equilibrium state is governed by the second smallest eigenvalue of the graph Laplacian $(\lambda_2)$ \cite{Olfati2004}.
Larger values of $\lambda_{2}$ results in faster convergence rate.
The CTC algorithm is also known as the continuous time Markov chain algorithm.
The inverse of the second smallest eigenvalue of the graph Laplacian $(\lambda_2)$ is referred to as the relaxation time \cite{KijimaBook1997}.

For a given connected network with an underlying graph topology $\mathcal{G}$, the Fastest Continuous Time Consensus (FCTC) problem can be formulated as below,
\begin{equation}
    \label{eq:FCTCOptmizationProblem1}
    \begin{aligned}
        \max\limits_{\boldsymbol{w}} \quad &\lambda_2(L_{\mathcal{G}}) \\
        s.t. \quad &\sum_{\{j,k\} \in \mathcal{E}} {\boldsymbol{w}_{jk}} \leq D,
    \end{aligned}
\end{equation}
where $\boldsymbol{w}_{jk}$ is the weight on the edge from node $j$ to node $k$ and $D$ is an upper limit on the total amount of weights.
This optimization problem can be defined in the form of standard semidefinite programming (see Appendix \ref{sec:SDP}) 
as below
\begin{equation}
    \nonumber
    \begin{aligned}
        \min\limits_{\boldsymbol{w}} \quad &-s \\
        s.t. \quad &L_{\mathcal{G}} + (-s)\boldsymbol{I} - (\boldsymbol{1} \times \boldsymbol{1}^{T}) / N  \succeq  \boldsymbol{0},  \\
                   & D - \sum_{\{j,k\} \in \mathcal{E}} {\boldsymbol{w}_{jk}}  \geq 0.
    \end{aligned}
\end{equation}
In the formulation above $\boldsymbol{1}$ is the column vector of all one.

To the best of our knowledge, analytical optimization of the CTC problem has been addressed only for tree topologies in \cite{Fiedler1990} by algebraic method.
In section \ref{sec:Lambda2SDP}, we have provided analytical solution to the semidefinite programming formulation of the CTC problem for a wider range of topologies.

An automorphism of the graph $\mathcal{G} = (\mathcal{V}, \mathcal{E})$ is a permutation $\sigma$ of $\mathcal{V}$ such that $\{i,j\} \in \mathcal{E}$ if and only if $\{\sigma(i),\sigma(j)\}\in \mathcal{E}$, the set of all such permutations, with composition as the group operation, is called the automorphism group of the graph and denoted by $Aut(\mathcal{G})$.
For a vertex $i \in \mathcal{V}$, the set of all images $\sigma(i)$, as $\sigma$ varies through a subgroup $G \subseteq Aut(\mathcal{G})$, is called the orbit of $i$ under the action of $G$.
The vertex set $\mathcal{V}$ can be written as disjoint union of distinct orbits.
In \cite{Ghosh06Boyd}, it has been shown that the optimal weights on the edges within an orbit are equal.

\section{Continuous Time Quantum Consensus}
\label{sec:ContinuousTimeQuantumConsensus}

\subsection{Lindblad Master Equation}
\label{sec:LindbladMasterEquation}
We consider a quantum network as a composite (or multipartite) quantum system with $N$ qudits.
Assuming $\mathcal{H}$ as the d-dimensional Hilbert space over $\mathbb{C}$, then the state space of the quantum network is within the Hilbert space $\mathcal{H}^{\otimes N} = \mathcal{H} \otimes \ldots \otimes \mathcal{H}$.
The state of the quantum system is described by its density matrix $(\boldsymbol{\rho})$.
This matrix is positive Hermitian and its trace is one $(tr(\boldsymbol{\rho}) = 1)$.
The network is associated with an underlying graph $\mathcal{G}=\{ \mathcal{V}, \mathcal{E} \}$, where $\mathcal{V}=\{1,\ldots, N\}$ is the set of indices for the $N$ qudits, and each element in $\mathcal{E}$ is an unordered pair of two distinct qudits, denoted as $\{j,k\} \in \mathcal{E}$ with $j,k \in \mathcal{V}$.
Permutation group $S_{N}$ acts in a natural way on $\mathcal{V}$ by mapping $\mathcal{V}$ onto itself.
For each permutation $\pi \in S_{N}$
we associate unitary operator $U_{\pi}$ over $\mathcal{H}^{\otimes N}$, as below
\begin{equation}
    \nonumber
    \begin{gathered}
        U_{\pi} ( Q_{1} \otimes \cdots \otimes Q_{N} ) = Q_{\pi(1)} \otimes \cdots \otimes Q_{\pi(N)},
     \end{gathered}
\end{equation}
where $Q_{i}$ is an operator in $\mathcal{H}$ for all $i = 1, \ldots, N$.
A special case of permutations is the swapping permutation or transposition where $\pi(j)=k$, $\pi(k)=j$ and $\pi(i) = i$ for all $i \in \mathcal{V}$ and $i \notin {j,k}$
We denote the swapping permutation between the qudits indices $j$ and $k$ by $\pi_{j,k}$ and the corresponding swapping operator by $U_{j,k}$.
In Appendix \ref{sec:GellMannMatrices} the swapping operator $U_{j,k}$ has been expressed as linear combination of the Cartezian product of Gell-Mann matrices.

Employing the quantum gossip interaction introduced in \cite{PetersenRef15}, the evolution of the quantum network can be described by the following master equation
\begin{equation}
    \label{eq:Lindblad2}
    \begin{gathered}
        \frac{d\boldsymbol{\rho}}{dt} = - \frac{i}{\hbar} [ H , \boldsymbol{\rho} ]  +  \sum_{ \{ j , k \} \in \mathcal{E} } { w_{j,k} \left(  U_{jk} \times \boldsymbol{\rho} \times U_{jk}^{\dagger} - \boldsymbol{\rho}  \right) }
     \end{gathered}
\end{equation}
where $w_{jk}$ is the positive constant weight over the edge ${j,k}$.
These weights form the distribution of limited amount of weight up to $D$, among edges of the underlying graph, i.e.
\begin{equation}
    \label{eq:Lindbladconstraint}
    \begin{gathered}
        \sum_{ \{j,k\} \in \mathcal{E} } {w_{j,k}} \leq D.
     \end{gathered}
\end{equation}
In order to have the set of transpositions corresponding to the edges of the underlying graph as the generator set $\mathcal{S}$ of the symmetric group $S_N$, the underlying graph should be connected.

For the case of $H = 0$, the evolution of $\boldsymbol{\rho}(t)$ is described in the following Lindblad master equation
\begin{equation}
    \label{eq:Lindblad3}
    \begin{gathered}
        \frac{d\boldsymbol{\rho}}{dt} = \sum_{ \{ j , k \} \in \mathcal{E} } { w_{j,k} \left(  U_{jk} \times \boldsymbol{\rho} \times U_{jk}^{\dagger} - \boldsymbol{\rho}  \right) }.
     \end{gathered}
\end{equation}
which is named Quantum Consensus Master Equation (QCME) by authors in \cite{Petersen2015IEEETranAutControl}, 
and its resultant quantum consensus state \cite{PetersenRef15} is defined as
\begin{equation}
    \label{eq:QCMEFinalConsensus}
    \begin{gathered}
        \boldsymbol{\rho}^{*}  =  \frac{1}{N!} \sum_{\pi \in S_{N}} {U_{\pi} \boldsymbol{\rho}(0) U_{\pi}^{\dagger} }.
     \end{gathered}
\end{equation}
In \cite{Petersen2015IEEETranAutControl}, it is shown that the QMCE reaches quantum consensus, namely $\lim_{t \rightarrow \infty} {\boldsymbol{\rho}}(t)  =  \boldsymbol{\rho}^{*}$ provided that the underlying graph of the quantum network is connected.

The aim of the analysis presented in the rest of this paper is to evaluate and optimize the convergence rate of the QCME to its quantum consensus state.
To this aim, we expand the density matrix $(\boldsymbol{\rho})$ as the linear combination of the generalized Gell-Mann matrices (introduced in Appendix \ref{sec:GellMannMatrices}) as below,
\begin{equation}
    \label{eq:DecompositionDensityGeneral}
    \begin{gathered}
        \boldsymbol{\rho} = \frac{1}{2^N} \sum_{ \mu_{1}, \mu_{2}, \ldots, \mu_{N} = 0 }^{ d^{2} - 1 }  {  \rho_{ \mu_{1}, \mu_{2}, \ldots, \mu_{N} } \cdot \lambda_{\mu_{1}} \otimes  \lambda_{\mu_{2}} \otimes \cdots \otimes \lambda_{\mu_{N}}  },
     \end{gathered}
\end{equation}
where $N$ is the number of particles and $\otimes$ denotes the Cartesian product and $\lambda$ matrices are the generalized Gell-Mann matrices as in (\ref{eq:DecompositionLambdaMatrices}) and (\ref{eq:DecompositionLambda0}).
Note that due to Hermity of density matrix, its coefficients of expansion $\rho_{ \mu_{1}, \mu_{2}, \ldots, \mu_{N} }$ are real numbers and because of unit trace of $\boldsymbol{\rho}$ we have $\rho_{0,0,\ldots,0} = 1$.

Using the decomposition of $\boldsymbol{\rho}$ in (\ref{eq:DecompositionDensityGeneral}), its permutations can be written as below
\begin{equation}
    \label{eq:DecompositionDensityPermutation}
    \begin{gathered}
        U_{j,k} \times \boldsymbol{\rho} \times U_{j,k}^{\dagger} = \frac{1}{2^N} \sum_{ \mu_{1}, \mu_{2}, \ldots \mu_{N} = 0 }^{ d^{2} - 1 }  {  \rho_{ \mu_{1}, \ldots \mu_{k}, \ldots, \mu_{j}, \ldots, \mu_{N} } \cdot \lambda_{\mu_{1}} \otimes \cdots \lambda_{\mu_{j}} \otimes \cdots \lambda_{\mu_{k}} \otimes \cdots \otimes \lambda_{\mu_{N}}  }
     \end{gathered}
\end{equation}
Note that in (\ref{eq:DecompositionDensityPermutation}) due to permutation operators, the place of indices $\mu_{j}$ and $\mu_{k}$ in the index of parameter $\boldsymbol{\rho}$ are interchanged.
Substituting the density matrix $\boldsymbol{\rho}$ from (\ref{eq:DecompositionDensityGeneral}) and its permutation (\ref{eq:DecompositionDensityPermutation}) in Lindblad master equation (\ref{eq:Lindblad3}) and considering the independence of the matrices $\lambda_{\mu_{1}} \otimes \lambda_{\mu_{2}} \otimes \cdots \lambda_{\mu_{N}} $ we can conclude the following for Lindblad master equation (\ref{eq:Lindblad3}),
\begin{equation}
    \label{eq:DensityEquation1}
    \begin{aligned}
        &\frac{d}{dt} \rho_{\mu_{1}, \cdots, 
        \mu_{N}}  =    \\
        &\quad\quad\quad\quad\quad\quad\quad\quad  \sum_{\{j,k\} \in \mathcal{E} }   {  w_{j,k} \left( \rho_{\mu_{1},\cdots,\mu_{k},\cdots,\mu_{j},\cdots,\mu_{N} }  -  \rho_{\mu_{1},\cdots,\mu_{j},\cdots,\mu_{k},\cdots,\mu_{N} }  \right)  } \\
        &\quad\quad\quad\quad\quad\quad\quad\quad\quad\quad\quad\quad\quad\quad\quad\quad\quad  \text{for all     } \mu_{1},\mu_{2},\cdots,\mu_{N}=0,\cdots,d^{2}-1,
    \end{aligned}
\end{equation}
with the constraint (\ref{eq:Lindbladconstraint}) on the edge weights.
Following the same procedure, the tensor component of the quantum consensus state (\ref{eq:QCMEFinalConsensus}) can be written as below
\begin{equation}
    \label{eq:QuantumConsensusState872}
    \begin{gathered}
        \rho_{\mu_1, \mu_2, \ldots, \mu_N}^{*}  =  \frac{1}{N!} \sum_{\pi \in S_{N}} {\rho_{\pi(\mu_1), \pi(\mu_2), \ldots, \pi(\mu_N)} (0)}
    \end{gathered}
\end{equation}
and for the connected underlying graph, the QCME reaches quantum consensus, componentwise as below
\begin{equation}
    \nonumber
    \begin{gathered}
        \lim_{t \rightarrow \infty} {    \rho_{\mu_1, \mu_2, \ldots, \mu_N}  (t)    }  =  \rho_{\mu_1, \mu_2, \ldots, \mu_N}^{*},
    \end{gathered}
\end{equation}

Comparing the set of equations in (\ref{eq:DensityEquation1}) with those of the CTC problem in (\ref{eq:CTCStateUpdate}) we can
see that the Quantum Consensus Master Equation (\ref{eq:Lindblad3}) is transformed into the classical CTC problem  (\ref{eq:CTCStateUpdate}) with $d^{2N} -1$ tensor component $\rho_{\mu_{1}, \cdots, \mu_{N}}$ as the agents' states.
Defining $\boldsymbol{X}_Q$ as a column vector of length $d^{2N}$ with components $\rho_{\mu_1, \ldots, \mu_N}$, the state update equation of the classical CTC can be written as below,
\begin{equation}
    \label{eq:QuantumStateUpdate}
    \begin{gathered}
        \frac{d\boldsymbol{X}_Q}{dt} = - \boldsymbol{L}_Q \boldsymbol{X}_Q.
    \end{gathered}
\end{equation}
$\boldsymbol{L}_Q$ is the corresponding Laplacian matrix as below,
\begin{equation}
    \label{eq:QuantumLaplacian}
    \begin{gathered}
        \boldsymbol{L}_Q  =  \sum_{\{j,k\} \in \mathcal{E}} {  \boldsymbol{w}_{j,k} ( I_{d^{2N}}  -  U_{j,k} )  },
    \end{gathered}
\end{equation}
where $U_{j,k}$ is the swapping operator given in Appendix \ref{sec:GellMannMatrices} (\ref{eq:PermutationGellMann}), 
 provided that $d$ is replaced with $d^2$ which in turn results in Gell-Mann matrices of size  $d^2 \times d^2$.
As explained in section \ref{sec:CTCPriliminaries}, the convergence rate of the obtained CTC problem is dictated by the spectral gap of its associated underlying graph which is the second largest eigenvalue $(\lambda_2 ( \boldsymbol{L}_Q ))$ of its Laplacian matrix $\boldsymbol{L}_Q$.
Thus, the corresponding Fastest Continuous Time Consensus problem can be written as the following optimization problem,
\begin{equation}
    \label{eq:FCTQCInitial}
    \begin{aligned}
        \max\limits_{\boldsymbol{w}} \quad &\lambda_2(\boldsymbol{L}_{Q}) \\
        s.t. \quad &\sum_{\{j,k\} \in \mathcal{E}} {\boldsymbol{w}_{jk}} \leq D.
    \end{aligned}
\end{equation}
We refer to this problem as the Fastest Continuous Time Quantum Consensus (FCTQC) problem.
This is the same optimization problem as in \cite{Petersen2015IEEETranAutControl} with the difference that in \cite{Petersen2015IEEETranAutControl} it has been obtained in computational basis.

The QCME (\ref{eq:Lindblad3}) reaches quantum consensus (\ref{eq:QCMEFinalConsensus}), due to the fact that the generating set is selected in a way that the whole group of $S_N$ can be generated, and the resultant Cayley graph of $S_N$ is connected.
Even though, the quantum consensus is achieved but surprisingly, the equations in (\ref{eq:DensityEquation1}) indicate that all agents are not able to exchange information with each other.
This is due to the fact that the underlying graph of the CTC problem obtained from (\ref{eq:DensityEquation1}) is not connected and the consensus is not reachable in the same sense as in the classical CTC problem, where the sufficient condition for reaching consensus is the connected underlying communication graph.

From the right hand side of the equation (\ref{eq:DensityEquation1}), we can see that the tensor components $\rho_{\mu_{1}, \cdots, \mu_{N}}$ that can be transformed into each other by permuting their indices are communicating and exchanging information with each other.
These tensor components or agent states correspond to Young tabloids of the same partition that in turn are equivalent to the agents in the classical CTC problem.
Thus, the agents belonging to the same partition form the connected components of the underlying graph of the classical CTC problem (\ref{eq:DensityEquation1}).

As mentioned above the underlying graph is a cluster of connected components where each connected graph component corresponds to a given partition of $N$ into $K$ integers, namely $N = n_{1} + n_{2} + \cdots + n_{K}$, where $K \leq d^2$ and $n_{j}$ for $j=1,\ldots,K$  is the number of indices in $\rho_{\mu_1, \mu_2, \ldots, \mu_N}$ with equal values. 
For a given partition and its associated Young Tabloids, more than one connected component can be obtained depending on the value of the $\mu$ indices.
As an example consider a quantum network with three qubits and the path graph as its underlying graph.
In this network, the values that the $\mu$ indices can take are $0, 1, 2$ and $3$.
For partition $n=(2,1)$ and Young Tabloids $t_n(1,1,2), t_n(1,2,1), t_n(2,1,1)$ and $\mu_1 = 0$  and $\mu_2 = 1$ the obtained underlying graph of the CTC problem is a path graph with three vertices where each vertex corresponds to one of the Young Tabloids mentioned above.
Now for the same partition and Young Tabloids but different values of the $\mu$ indices (e.g. $\mu_1 = 1$  and $\mu_2 = 0$) the obtained underlying graph of the CTC problem is same as that of the previous example.
As a matter of fact for this partition, there are $12$ connected components which are identical to each other.
Each one of these connected components has an identical impact on the convergence rate of the QCME to its quantum consensus state.
Therefore for each partition we consider only one of them, and we refer to this graph as the induced graph.
The only exception is the case of $N=d^2$ where there is only one connected component corresponding to the partition that all indices are different from each other.
These induced graphs are the same as those noted in \cite{Petersen2015IEEETranAutControl}.

For the given partition $n=[n_1, n_2, \cdots, n_K]$, using the Yamanouchi symbol (introduced in section \ref{sec:YoungHasse}) a Young tabloid of partition $n$ is uniquely represented by the notation $t_{n}(r_{1}, r_{2}, \cdots, r_{N-1}, r_{N})$.
Each Young tabloid $t_{n}(r_{1}, r_{2}, \cdots, r_{N})$ is equivalent to an agent in the 
induced graph of the CTC problem
and its corresponding coefficient ($\rho_{\mu_{r_{1}}, \mu_{r_{2}}, \cdots, \mu_{r_{N}}}$) is equivalent to the state of that agent.

The CTC equation obtained from (\ref{eq:DensityEquation1}) for partition $n$ is as below,
\begin{equation}
    \label{eq:Interlacing1}
    \begin{aligned}
        \frac{d}{dt} &\rho_{   \mu_{r_{1}(m)},  \ldots, \mu_{r_{N}(m)}   }
        =  \quad\quad\quad\quad\quad\quad\quad\quad \\
        &\sum_{\{j,l\} \in \mathcal{E} }  {}   w_{j,l} \cdot \left(   \rho_{\mu_{\pi_{j,l} (r_{1}(m))},\ldots, \mu_{\pi_{j,l} (r_{N}(m))}}    \right.
        -  \\
        &   \quad \quad \quad \quad \quad \quad \left.
        \rho_{   \mu_{r_{1}(m)},  \ldots, \mu_{r_{N}(m)}   }
        \right),
    \end{aligned}
\end{equation}
where $m$ varies from $1$ to $\nu = N! / (n_1! \cdot n_2! \cdots n_K!)$
and $\pi_{j,l}$ transposes the $j$-th and $l$-th Yamanouchi symbols i.e. $r_j$ and $r_l$.
Note that for the agent states that their Yamanouchi symbols $(r_{j}, r_{l})$ are equal, the value inside the summation above is zero.

We define the column vector $\boldsymbol{X}_{n}$ as the state vector of the associated CTC problem (\ref{eq:CTC12NewFormat}) of agiven partition $n$.
This vector includes the tensor components corresponding to the Young Tabloids of the partition $n$ and it has ${\nu = N!/(n_1! \cdot n_2! \cdots n_K!)}$ elements.
As mentioned above the underlying graph of the CTC problem is a cluster of connected components,
i.e. the Laplacian matrix $\boldsymbol{L}_Q$ is a block diagonal matrix where each block corresponds to one of the connected components,
with  state vector  $\boldsymbol{X}_{n}$.
The state update equation (\ref{eq:QuantumStateUpdate}) for the state vector $\boldsymbol{X}_{n}$ is as below,
\begin{equation}
    \label{eq:StateUpdateEquationXn}
    \begin{gathered}
        \frac{d\boldsymbol{X}_n}{dt} = -\boldsymbol{L}_n \boldsymbol{X}_n,
    \end{gathered}
\end{equation}
with $\boldsymbol{L}_n$ as the Laplacian matrix which is one of the blocks in $\boldsymbol{L}_Q$.

The tensor component of the quantum consensus state (\ref{eq:QuantumConsensusState872}) for partition $n$ takes the following form
\begin{equation}
    \label{eq:QuantumConsensusState13}
    \begin{gathered}
        \rho_{   \mu_{r_{1}},  \ldots, \mu_{r_{N}}   }^{*}   =  \frac{1}{N!} \sum_{\pi \in S_{N}} { \rho_{ \mu_{\pi(r_{1})},  \ldots, \mu_{\pi(r_{N})} } }
    \end{gathered}
\end{equation}

As explained in section \ref{sec:YoungHasse}, $S_N$ acts transitively over the set of Young tabloids or agents and consequently over the following set of agent states
$$(\{  \{ \rho_n (r_1(1), r_2(1), \cdots, r_N(1)) \},  \cdots,  \{ \rho_n (r_1(\nu), r_2(\nu), \cdots, r_N(\nu)) \}  \})$$
with the Young subgroup $S_{n}$ as its stabilizer subgroup.
Since the group elements of the Young subgroup do not change the Yamanouchi symbols.
Based on the one to one correspondence between agent states and the right or left cosets of $S_n$ in $S_N$, it can be concluded that the connected component is the Schreier coset graph of permutation group $S_{N}$  with Young subgroup $S_{n}$ and generating set consisting of transpositions associated with edges of the underlying graph of the quantum network.
For the case of trivial $S_n$ (i.e. $n=[1, 1, \ldots, 1]$) the Schreier coset graph is reduced to Cayley graph.

In the following, we provide some of the typical partitions with their corresponding connected graph component.

The most simple case is the one that all indices are the same, i.e. the partition is $n= (n_{1})=(N)$.
The Yamanouchi symbols for this partition are $r_{1} = r_{2} = \cdots = r_{N} = 1$ and the CTC equation (\ref{eq:Interlacing1}) is as below
\begin{equation}
    \nonumber
    \begin{aligned}
        \frac{d}{dt} \rho_{\mu_1,\mu_1,\ldots,\mu_1} = 0, \quad\quad \text{for} \quad\quad \mu_1=0,1,\ldots,d^2-1.
    \end{aligned}
\end{equation}
and therefore
\begin{equation}
    \nonumber
    \begin{aligned}
        \rho_{\mu_1, \mu_1, \ldots,\mu_1}(t) = \rho_{\mu_1, \mu_1, \ldots, \mu_1}(0), \quad\quad \text{for} \quad\quad \mu_1=0,1,\ldots,d^2-1.
    \end{aligned}
\end{equation}
The induced graph of this partition is the edgeless or the empty graph that is a graph without any edges.
This is the Schreier coset graph $Sch(S_N, \mathcal{S}, S_N)$.
Due to lack of any information exchange between agents, the agent states does not change by time, and they maintain their initial values.
Thus for the quantum consensus state (\ref{eq:QuantumConsensusState13}) we have
\begin{equation}
    \nonumber
    \begin{aligned}
        \rho_{\mu_1, \mu_1, \ldots,\mu_1}^{*} = \frac{1}{N!} \sum_{\pi \in S_{N}} { \rho_{\pi(\mu_1), \pi(\mu_1), \ldots, \pi(\mu_1)}(0) }
        = \rho_{\mu_1, \mu_1, \ldots, \mu_1}(0),
    \end{aligned}
\end{equation}
where the second equality above is based on the fact that the agent state $\rho_{\mu_1, \mu_1, \ldots,\mu_1}$ remains intact under the permutation $\pi$ or any exchange of information.

The next non-trivial partition is the case where all $\mu$ indices are the same except for one of them, thus we have
\begin{equation}
    \nonumber
    \begin{aligned}
        n = [N-1, 1],
    \end{aligned}
\end{equation}
i.e. the Yamanouchi symbols are as $r_{i} = 1$ for $i=\{1,...,N\} \setminus \{j\}$  and $r_{j} = 2$.
Thus the agent state can be written as $\rho_{\mu_1,\ldots,\mu_1,\mu_{2},\mu_1,\ldots,\mu_1}$ where for ease of notation we denote the agent state by the scalar variable $x_{j}$ for $j=1, \ldots, N$.
Hence the CTC equation (\ref{eq:Interlacing1}) for the partition $n=[N-1,1]$  can be written as below,
\begin{equation}
    \label{eq:ContinuousConsensusMain}
    \begin{aligned}
        \frac{d}{dt} x_{j} = \sum_{k \in \mathcal{N}(j)}  { w_{j,k} ( x_{k} - x_{j} ) },
    \end{aligned}
\end{equation}
with the constraint (\ref{eq:Lindbladconstraint}) on the edge weights.
$\mathcal{N}(j)$ is the set of neighbours of node $j$ in the graph $\mathcal{G}$.
Considering $x_{j}$ as the state for node $j$, the equation above is same as the classical CTC problem over the underlying graph $\mathcal{G}$
which in turn is the Schreier graph $Sch(S_N, \mathcal{S}, S_{N-1})$.
For this particular partition, the induced graph of the partition is same as the underlying graph of the quantum network $(\mathcal{G})$.
For the quantum consensus state (\ref{eq:QuantumConsensusState872}) of this partition we have
\begin{equation}
    \nonumber
    \begin{gathered}
        \rho_{\mu_1, \mu_1, \ldots,\mu_1, \mu_2}^{*} =
        \rho_{\mu_1, \mu_1, \ldots,\mu_1, \mu_2, \mu_1}^{*} =
        \cdots  =
        \rho_{\mu_2, \mu_1, \ldots,\mu_1, \mu_1}^{*} =
        \\
         \frac{1}{N} \sum_{j=1}^{N} { \rho_{\mu_1, \ldots, \mu_1, \underbrace{\scriptstyle\mu_2}_{j\text{-th}}, \mu_1, \ldots, \mu_1}(0) }
        = \frac{1}{N} \sum_{j=1}^{N} {x_{j}(0)},
    \end{gathered}
\end{equation}
Note that, in this case, the quantum consensus state is same as the final equilibrium state of the classical CTC problem.

For the case that all indices are different, namely for the partition $n = [1,1,\ldots,1]$, the CTC problem is referred to as interchange Process \cite{ProofAldous}.
This case is possible if $N \leq d^2$.
The Yamanouchi symbols for this partition take different values from $1$ to $N$ where no two symbols are equal to each other.
The quantum consensus state (\ref{eq:QuantumConsensusState872}) for this partition is same as (\ref{eq:QuantumConsensusState13}) with the exception that no two $\mu$ indices have the same value.
The induced graph of this partition is the Schreier coset graph $Sch(S_N, \mathcal{S}, e)$ where $e$ is the identity element of $S_N$.
This Schreier coset graph is same as the Cayley graph $(S_N, \mathcal{S})$.

\begin{figure}
  \centering
     \includegraphics[width=130mm]{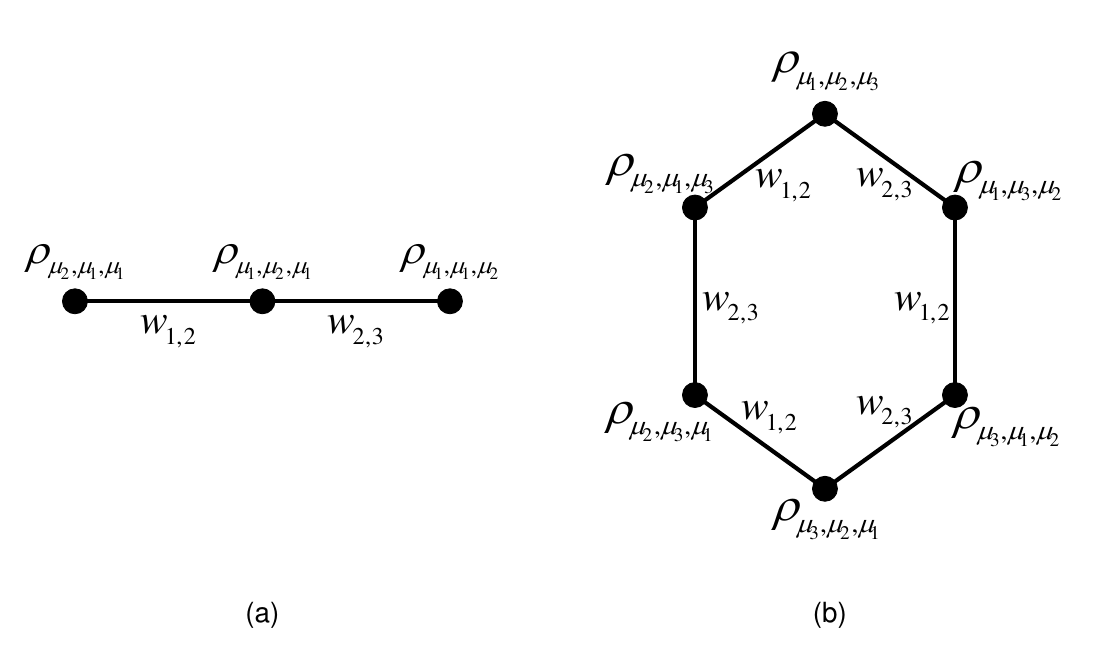}
  \caption{The induced graphs for a path graph with 3 vertices for partitions (a) $n=(2,1)$ and (b) $n=(1,1,1)$.}
  \label{fig:InducedGraphs4Path3}
\end{figure}

As an example, consider the path graph with three vertices (denoted by $\mathcal{G}_{P3}$) as the underlying graph of the quantum network.
For partition $n=(2,1)$ over graph $\mathcal{G}_{P3}$ the induced graph 
is as depicted in figure \ref{fig:InducedGraphs4Path3} (a) which is same as the underlying graph $\mathcal{G}_{P3}$.
But for partition $n=(1,1,1)$ the induced graph obtained is a cycle graph as depicted in figure \ref{fig:InducedGraphs4Path3} (b).

\subsection{ Intertwining of Induced Graph }

As stated before, the aim of this paper is to evaluate and optimize the convergence rate of the QCME (\ref{eq:Lindblad3}) to its quantum consensus state (\ref{eq:QCMEFinalConsensus}).
In doing so in section \ref{sec:LindbladMasterEquation}, we have shown that the QCME (\ref{eq:Lindblad3}) can be modeled as a classical CTC problem with a disconnected underlying graph that is a cluster of connected components referred to as the induced graphs.
Therefore, the convergence rate of the QCME (\ref{eq:Lindblad3}) is equivalent to the convergence rate of the obtained classical CTC problem,
i.e. the second largest eigenvalue of its Laplacian matrix $(\lambda_2(\boldsymbol{L}_Q))$, which in turn is determined by the second largest eigenvalues of the induced graphs.
In this section, we address the order of the second largest eigenvalues of the induced graphs.
We have shown that the second eigenvalues of all induced graphs are equal and based on this we will show that the general problem reduces to finding the second largest eigenvalue of the induced graph of the least dominant partition in the Hasse diagram (as explained in section \ref{sec:YoungHasse}).
To this aim, we show that the eigenvalues of the induced graph corresponding to the dominant partition (higher level of the Hasse diagram) is included in the eigenvalues of the less dominant partition (lower level of the Hasse diagram).

For ease of notation we consider the Young Tabloid where the Yamanouchi symbols are sorted i.e. $r_{D_i + 1} = r_{D_i + 2} = \cdots = r_{D_i + n_i} = i$ with $D_i = \sum_{j=1}^{i-1}{n_j}$ and $D_1 = 0$.
It is obvious that the other Tabloids of this partition can be obtained from the permutation of Yamanouchi symbols of the above Young Tabloid.
Using this notation, the CTC equations (\ref{eq:Interlacing1}) for a given partition can be written as below
\begin{equation}
    \label{eq:CTC12NewFormat}
    \begin{aligned}
        \frac{d}{dt} &\rho_{   \underbrace{\scriptstyle\mu_{\pi(1)},\ldots,\mu_{\pi(1)}}_{n_{1}},  \ldots, \underbrace{\scriptstyle\mu_{\pi(K)},\ldots,\mu_{\pi(K)}}_{n_{K}}   }
        =  \quad\quad\quad\quad\quad\quad\quad\quad \\
        &\sum_{\{j,l\} \in \mathcal{E} }  {}   w_{j,l} \cdot \left(   \rho_{\underbrace{\scriptstyle\mu_{\pi_{j,l}\pi(1)},\ldots,\mu_{\pi_{j,l}\pi(1)}}_{n_{1}},
        \ldots,
        \underbrace{\scriptstyle\mu_{\pi_{j,l}\pi(K)},\ldots,\mu_{\pi_{j,l}\pi(K)}}_{n_{K}}  }  \right.
        -  \\
        &   \quad \quad \quad \quad \quad \quad \left.
        \rho_{\underbrace{\scriptstyle\mu_{\pi(1)},\ldots,\mu_{\pi(1)}}_{n_{1}},  \ldots, \underbrace{\scriptstyle\mu_{\pi(K)},\ldots,\mu_{\pi(K)}}_{n_{K}} }  \right)
    \end{aligned}
\end{equation}
where $\pi \in S_N$ permutes the location of indices and $\pi_{j,l}$ transposes the location of $j$-th and $l$-th indices.

In Hasse diagram, the one level dominance i.e. the partitions that are one level apart in Hasse diagram can be classified into two categories.
First category is the case when one of the boxes in the Young diagram is displaced from a higher row to an existing lower one in the Young diagram, provided that the new diagram is again a Young diagram.
An example for this category is the partitions $(4,2)$ and $(3,3)$ in the Hasse diagram depicted in figure \ref{fig:HasseDiagram}.
The second category is the case when one of the indices is changed
to a new value.
In terms of Young diagram of the partitions, one box is moved from a higher row to a new row at the bottom of the Young diagram.
An example for this category is the partitions $(4,2)$ and $(4,1,1)$ in the Hasse diagram depicted in figure \ref{fig:HasseDiagram}.

\subsubsection*{First Category}
Consider two given partitions of $N$ namely, $n$ and $n^{'}$, where partition $n$ is one level dominant to partition $n^{'}$.
If the dominance level is of first category then
Partition $n^{'}$ can be written in terms of partition $n = [n_{1}, n_{2}, \ldots, n_{K}]$ as following $n^{'} = [ n_{1}, n_{2}, \ldots, n_{m}-1, n_{m+1}, \ldots, n_{r-1}, n_{r}+1, \ldots, n_{K} ]$,
provided that $n_{m} > n_{m+1}$ and $n_{r} < n_{r-1}$.
Considering the derivation of the CTC equation (\ref{eq:CTC12NewFormat}) for partition $n^{'}$, we can define the new variable $\tilde{\rho}$ in terms of tensor components $\rho$ as below
\begin{equation}
    \label{eq:InterlacingATildaSecondCase}
    \begin{aligned}
        &\tilde{\rho}_{   \underbrace{\scriptstyle\mu_{\pi(1)},\ldots,\mu_{\pi(1)}}_{n_{1}},
        \ldots,
        \underbrace{\scriptstyle\mu_{\pi(m)},\ldots,\mu_{\pi(m)}}_{n_{m}},
        \ldots,
        \underbrace{\scriptstyle\mu_{\pi(r)},\ldots,\mu_{\pi(r)}}_{n_{r}},
        \ldots,
        \underbrace{\scriptstyle\mu_{\pi(K)},\ldots,\mu_{\pi(K)}}_{n_{K}}
        }
        = \\
        & 
        \rho_{   \underbrace{\scriptstyle\mu_{\pi(1)},\ldots,\mu_{\pi(1)}}_{n_{1}},
        \ldots,
        \underbrace{\scriptstyle\mu_{\pi(r)},\mu_{\pi(m)},\ldots,\mu_{\pi(m)}}_{n_{m}},
        \ldots,
        \underbrace{\scriptstyle\mu_{\pi(r)},\mu_{\pi(r)},\ldots,\mu_{\pi(r)}}_{n_{r}},
        \ldots,
        \underbrace{\scriptstyle\mu_{\pi(K)},\ldots,\mu_{\pi(K)}}_{n_{K}}
        }
        \\
        &  
        +  \rho_{   \underbrace{\scriptstyle\mu_{\pi(1)},\ldots,\mu_{\pi(1)}}_{n_{1}},
        \ldots,
        \underbrace{\scriptstyle\mu_{\pi(m)},\mu_{\pi(r)},\mu_{\pi(m)},\ldots,\mu_{\pi(m)}}_{n_{m}},
        \ldots,
        \underbrace{\scriptstyle\mu_{\pi(r)},\ldots,\mu_{\pi(r)},\mu_{\pi(r)}}_{n_{r}},
        \ldots,
        \underbrace{\scriptstyle\mu_{\pi(K)},\ldots,\mu_{\pi(K)}}_{n_{K}}
        }
        \\
        &
        +  \cdots
        \\
        &
        +  \rho_{   \underbrace{\scriptstyle\mu_{\pi(1)},\ldots,\mu_{\pi(1)}}_{n_{1}},
        \ldots,
        \underbrace{\scriptstyle\mu_{\pi(m)},\ldots,\mu_{\pi(m)},\mu_{\pi(r)}}_{n_{m}},
        \ldots,
        \underbrace{\scriptstyle\mu_{\pi(r)},\ldots,\mu_{\pi(r)}}_{n_{r}},
        \ldots,
        \underbrace{\scriptstyle\mu_{\pi(K)},\ldots,\mu_{\pi(K)}}_{n_{K}}
        }
    \end{aligned}
\end{equation}
Taking the derivative of $\tilde{\rho}$ in (\ref{eq:InterlacingATildaSecondCase}) and applying the CTC equation of partition $n^{'}$ (\ref{eq:CTC12NewFormat}) to the right hand side of the resultant equation, it is straightforward to show that $\tilde{\rho}$ obeys the same CTC equations of partition $n$ as in (\ref{eq:CTC12NewFormat}).

\subsubsection*{Second Category}

If the dominance level is of second category then
Partition $n^{'}$ can be written in terms of partition $n = [n_{1}, n_{2}, \ldots, n_{K}]$ as $n^{'} = [n_{1}, \ldots, n_{m}-1, n_{m+1}, \ldots, n_{k}, 1 ]$,
provided that $n_{m} > n_{m+1}$.
Similar to first category, considering the derivation of the CTC equation (\ref{eq:CTC12NewFormat}) for partition $n^{'}$, we can define the new variable $\tilde{\rho}$ in terms of tensor components $\rho$ as below
\begin{equation}
    \label{eq:InterlacingATilda}
    \begin{aligned}
        &\tilde{\rho}_{   \underbrace{\scriptstyle\mu_{\pi(1)},\ldots,\mu_{\pi(1)}}_{n_{1}},  \ldots, \underbrace{\scriptstyle\mu_{\pi(m)},\ldots,\mu_{\pi(m)}}_{n_{m}},
        \ldots,
        \underbrace{\scriptstyle\mu_{\pi(K)},\ldots,\mu_{\pi(K)}}_{n_{K}}
        }
        = \\
        &
        \rho_{   \underbrace{\scriptstyle\mu_{\pi(1)},\ldots,\mu_{\pi(1)}}_{n_{1}},
        \ldots,
        \underbrace{\scriptstyle\mu_{\pi(m)},\ldots,\mu_{\pi(m)}}_{n_{m}-1},
        \ldots,
        \underbrace{\scriptstyle\mu_{\pi(K)},\ldots,\mu_{\pi(K)}}_{n_{K}}
        \underbrace{\scriptstyle\mu_{\pi(K+1)}}_{1}
        }
        \\
        &
        +  \rho_{   \underbrace{\scriptstyle\mu_{\pi(1)},\ldots,\mu_{\pi(1)}}_{n_{1}},
        \ldots,
        \underbrace{\scriptstyle\mu_{\pi(m)},\ldots,\mu_{\pi(m)},\mu_{\pi(K+1)}}_{n_{m}-1},
        \ldots,
        \underbrace{\scriptstyle\mu_{\pi(K)},\ldots,\mu_{\pi(K)}}_{n_{K}}
        \underbrace{\scriptstyle\mu_{\pi(m)}}_{1}
        }
        \\
        &
        +  \rho_{   \underbrace{\scriptstyle\mu_{\pi(1)},\ldots,\mu_{\pi(1)}}_{n_{1}},
        \ldots,
        \underbrace{\scriptstyle\mu_{\pi(m)},\ldots,\mu_{\pi(m)},\mu_{\pi(K+1)},\mu_{\pi(m)}}_{n_{m}-1},
        \ldots,
        \underbrace{\scriptstyle\mu_{\pi(K)},\ldots,\mu_{\pi(K)}}_{n_{K}}
        \underbrace{\scriptstyle\mu_{\pi(m)}}_{1}
        }
        \\
        &
        +  \cdots
        \\
        &
        +  \rho_{   \underbrace{\scriptstyle\mu_{\pi(1)},\ldots,\mu_{\pi(1)}}_{n_{1}},
        \ldots,
        \underbrace{\scriptstyle\mu_{\pi(K+1)},\mu_{\pi(m)},\ldots,\mu_{\pi(m)}}_{n_{m}-1},
        \ldots,
        \underbrace{\scriptstyle\mu_{\pi(K)},\ldots,\mu_{\pi(K)}}_{n_{K}}
        \underbrace{\scriptstyle\mu_{\pi(m)}}_{1}
        }.
    \end{aligned}
\end{equation}
In the same manner as in first category, after taking the derivative of $\tilde{\rho}$ in (\ref{eq:InterlacingATilda}) and applying the CTC equation of partition $n^{'}$ (\ref{eq:CTC12NewFormat}) to the right hand side of the resultant equation, it is straightforward to show that $\tilde{\rho}$ obeys the same CTC equations of partition $n$ as in (\ref{eq:CTC12NewFormat}).

For both categories, we have shown that the newly defined variable $\tilde{\rho}$ obeys the same CTC equations of partition $n$ as in (\ref{eq:CTC12NewFormat}).
Using this result, in the following we show that all eigenvalues of the partition $n$ are amongst the eigenvalues of the partition $n^{'}$.

we define the column vector $\tilde{\boldsymbol{X}}_n$ as the state vector for partition $n$ with $\tilde{\rho}$ as it components.
Depending on the category $\tilde{\rho}$ can be from the left hand side of either equations (\ref{eq:InterlacingATildaSecondCase}) or (\ref{eq:InterlacingATilda}).
Recalling $\boldsymbol{X}_{n^{'}}$ as the state vector of partition $n^{'}$ defined in section \ref{sec:LindbladMasterEquation} in terms of tensor components $\rho$, we can conclude the following, based on equations (\ref{eq:InterlacingATildaSecondCase}) and (\ref{eq:InterlacingATilda})
\begin{equation}
    \label{eq:XTildanXnPrime}
    \begin{aligned}
        \tilde{\boldsymbol{X}}_n   =   \boldsymbol{P}(n \rightarrow n^{'}) \times \boldsymbol{X}_{n^{'}}
    \end{aligned}
\end{equation}
where $\boldsymbol{P}(n \rightarrow n^{'})$ is the projection matrix with $\nu = N!/(n_1! \cdot n_2! \cdots n_K!)$ rows and $\nu^{'} = N!/(n^{'}_1! \cdot n^{'}_2! \cdots n^{'}_{K^{'}}!)$ columns.
Matrix $\boldsymbol{P}(n \rightarrow n^{'})$ is the projection matrix for the surjective projection that maps the states of partition $n^{'}$ onto sates of partition $n$.
Taking the derivative of the equation (\ref{eq:XTildanXnPrime}) and substituting the derivative of the state vectors with the products of state vectors and their associated Laplacian matrices according to state update equation (\ref{eq:StateUpdateEquationXn}) we can conclude that the following relation holds between the Laplacian matrices associated with partitions $n$ and $n^{'}$
\begin{equation}
    \label{eq:LplacianRelation}
    \begin{gathered}
        \boldsymbol{L}_n     \times     \boldsymbol{P}(n \rightarrow n^{'})
        =
        \boldsymbol{P}(n \rightarrow n^{'})     \times     \boldsymbol{L}_{n^{'}},
    \end{gathered}
\end{equation}
This is known as the intertwining relation.
By taking the transpose of both sides of this equation we obtain
\begin{equation}
    \label{eq:LplacianRelationTranspose}
    \begin{gathered}
        \boldsymbol{P}^{T}(n \rightarrow n^{'})     \times     \boldsymbol{L}_n
        =
        \boldsymbol{L}_{n^{'}}     \times     \boldsymbol{P}^{T}(n \rightarrow n^{'}),
    \end{gathered}
\end{equation}
Then for each eigenvalue of $\boldsymbol{L}_n$ denoted by $\gamma$ and its associated eigenvector $\boldsymbol{\Gamma}$ we have
\begin{equation}
    \nonumber
    \begin{gathered}
        \boldsymbol{L}_n    \times    \boldsymbol{\Gamma}
        =
        \gamma              \cdot     \boldsymbol{\Gamma}.
    \end{gathered}
\end{equation}
Multiplying both sides by $\boldsymbol{P}^{T}(n \rightarrow n^{'})$ from left
\begin{equation}
    \nonumber
    \begin{gathered}
        \boldsymbol{P}^{T}(n \rightarrow n^{'})     \times     \boldsymbol{L}_n    \times    \boldsymbol{\Gamma}
        =
        \gamma              \cdot     \left( \boldsymbol{P}^{T}(n \rightarrow n^{'})     \times     \boldsymbol{\Gamma} \right),
    \end{gathered}
\end{equation}
Using (\ref{eq:LplacianRelationTranspose}) we obtain the following
\begin{equation}
    \label{eq:LplacianRelationTranspose2}
    \begin{gathered}
        \boldsymbol{L}_{n^{'}}      \times      \left( \boldsymbol{P}^{T}(n \rightarrow n^{'})     \times    \boldsymbol{\Gamma} \right)
        =
        \gamma              \cdot     \left( \boldsymbol{P}^{T}(n \rightarrow n^{'})     \times     \boldsymbol{\Gamma} \right),
    \end{gathered}
\end{equation}
Since $\boldsymbol{P}(n \rightarrow n^{'})$ is the projection matrix for a surjective projection then $\boldsymbol{P}^{T}(n \rightarrow n^{'})$ is the projection matrix for an injective projection and it does not have a null space i.e. it doesn't have zero eigenvalue \cite{LinearAlgebraHoffman1971}.
Therefore,s it can be concluded that any  eigenvalue of $\boldsymbol{L}_n$ is also an eigenvalue of $\boldsymbol{L}_{n^{'}}$.

Based on this conclusion and the fact that the first eigenvalue $(\lambda_1)$ of both Laplacian matrices $\boldsymbol{L}_n$  and  $\boldsymbol{L}_{n^{'}}$ are zero then we can conclude the following relation for the second eigenvalue of the Laplacian matrices,
\begin{equation}
    \label{eq:LplacianSecondEigenvalueRelations}
    \begin{gathered}
        \lambda_2 (\boldsymbol{L}_{n^{'}})      \leq   \lambda_2 ( \boldsymbol{L}_n )
    \end{gathered}
\end{equation}
In simple words, we have shown that the second eigenvalue of a partition is less than or equal to that of its one level dominant partition.
Applying this conclusion to the partitions in the Hasse diagram of a given $N$, we can conclude that the second largest eigenvalue of the partition at bottom (top) of the Hasse diagram is the smallest (greatest) and the second largest eigenvalue of all other partition are in between with the order same as the dominance order in Hasse diagram.
In other words,
\begin{equation}
    \label{eq:LaplacianSecondEigenvalueHasseOrder}
    \begin{gathered}
        \lambda_2 ([\underbrace{\scriptstyle 1,1,\ldots,1}_{N}])      \leq   \lambda_2 ( [2, \underbrace{\scriptstyle 1,1,\ldots,1}_{N-2}] ) \leq \cdots \leq \lambda_2 ([N-1, 1])
    \end{gathered}
\end{equation}
Note that the partition $n=[N]$ has only on eigenvalue that is zero.

In the prominent work \cite{ProofAldous} authors have proved that the second eigenvalues (i.e. the spectral gap) of the partitions $[\underbrace{\scriptstyle 1,1,\ldots,1}_{N}]$ and $[N-1, 1]$ (known as the interchange and the random walk processes, respectively) are equal.
This is known as the Aldous' conjecture \cite{AldousBook}.
Considering this result and the relation (\ref{eq:LaplacianSecondEigenvalueHasseOrder}) it can be concluded that the second eigenvalues of all partitions (except $[N]$) in the Hasse diagram are equal to each other.
This is the generalization of the Aldous' conjecture to all partitions (except $[N]$) in the Hasse diagram of $N$.

An interesting point is a case where one of the induced graphs namely $\mathcal{G^{'}}$ with $N^{'}$ vertices serves as the underlying graph for a quantum network. Since the induced graph corresponding to partition $(N^{'}-1, 1)$ is same as the underlying graph $\mathcal{G^{'}}$, therefore, the spectral gap and the convergence rate of all induced graphs corresponding to partitions of $N^{'}$ are same as those of all induced graphs corresponding to partitions of $N$.

\begin{figure}
  \centering
     \includegraphics[width=100mm]{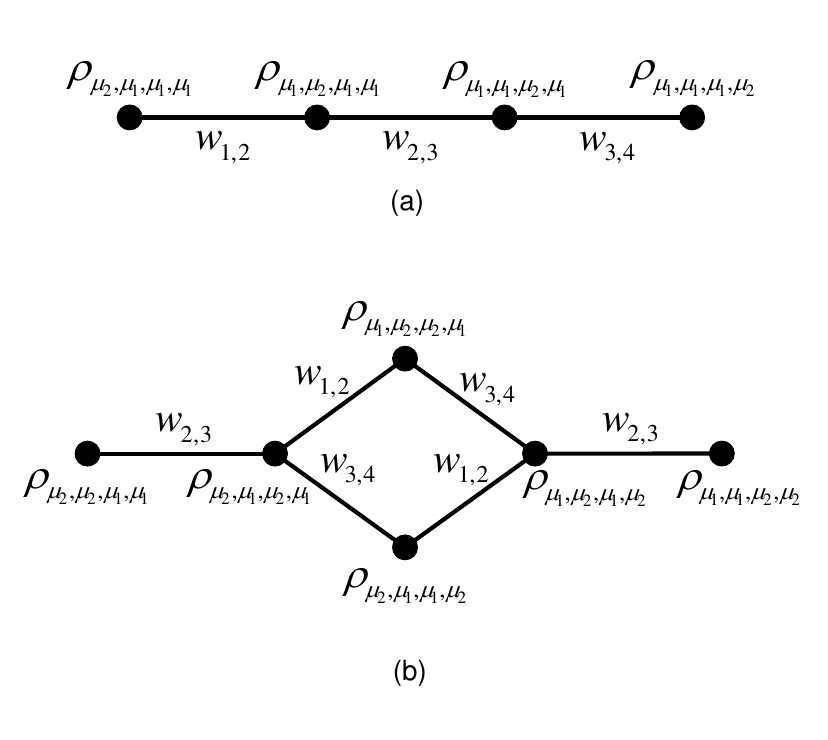}
  \caption{The induced graphs for a path graph with $4$ vertices for partitions (a) $n = (3, 1)$ and (b) $n = (2, 2)$.}
  \label{fig:InducedGraphs4Path4}
\end{figure}

In the following, we provide examples of the projection matrices $\boldsymbol{P}(n \rightarrow n^{'})$ for both categories of one level dominance as mentioned above.
For the first category, we consider the path graph with four vertices as the underlying graph of the quantum network.
The projection matrix from partition $n=[3,1]$ to partition $n^{'}=[2,2]$, and their Laplacian matrices are as below,
\begin{equation}
    \label{eq:ProjectionMatrixExampleCategory1}
        \boldsymbol{P}(n \rightarrow n^{'}) = \left[ \begin{array}{cccccc}
        {1} & {1} & {0} & {1} & {0} & {0}  \\
        {1} & {0} & {1} & {0} & {1} & {0}  \\
        {0} & {1} & {1} & {0} & {0} & {1}  \\
        {0} & {0} & {0} & {1} & {1} & {1}
        \end{array} \right],
\end{equation}
\begin{equation}
    \label{eq:Laplacian31ExampleCategory1}
        \boldsymbol{L}_n = \left[ \begin{array}{cccc}
        {w_{1,2}} & {-w_{1,2}} & {0} & {0}  \\
        {-w_{1,2}} & {w_{1,2}+w_{2,3}} & {-w_{2,3}} & {0}  \\
        {0} & {-w_{2,3}} & {w_{2,3}+w_{3,4}} & {-w_{3,4}}  \\
        {0} & {0} & {-w_{3,4}} & {w_{3,4}}
        \end{array} \right],
\end{equation}
\begin{localsize}{10}
\begin{equation}
    \label{eq:Laplacian22ExampleCategory1}
    \begin{aligned}
    &\boldsymbol{L}_{n^{'}} = \\
    &\left[ \begin{array}{cccccc}
        {w_{2,3}}   & {-w_{2,3}}                    & {0}                   & {0}           & {0}           & {0}  \\
        {-w_{2,3}}  & {w_{1,2}+w_{2,3}+w_{3,4}}     & {-w_{1,2}}            & {-w_{3,4}}    & {0}           & {0}  \\
        {0}         & {-w_{1,2}}                    & {w_{1,2}+w_{3,4}}     & {0}           & {-w_{3,4}}    & {0}  \\
        {0}         & {-w_{3,4}}                    & {0}                   & {w_{1,2}+w_{3,4}} & {-w_{1,2}} & {0}  \\
        {0} & {0}   & {-w_{3,4}} & {-w_{1,2}} & {w_{1,2}+w_{2,3}+w_{3,4}} & {-w_{2,3}}  \\
        {0} & {0} & {0} & {0} & {-w_{2,3}} & {w_{2,3}}
        \end{array} \right].
    \end{aligned}
\end{equation}
\end{localsize}
The induced graphs of these partitions are depicted in figure \ref{fig:InducedGraphs4Path4}.

As an example for the second category, consider the path graph with three vertices as the underlying graph of the quantum network.
The projection matrix from partition $n=[2,1]$ to partition $n^{'}=[1,1,1]$, and their Laplacian matrices are as below,
\begin{equation}
    \label{eq:ProjectionMatrixExampleCategory2}
        \boldsymbol{P}(n \rightarrow n^{'}) = \left[ \begin{array}{cccccc}
        {1} & {1} & {0} & {0} & {0} & {0}  \\
        {0} & {0} & {1} & {1} & {0} & {0}  \\
        {0} & {0} & {0} & {0} & {1} & {1}
        \end{array} \right],
\end{equation}
\begin{equation}
    \label{eq:Laplacian21ExampleCategory2}
        \boldsymbol{L}_n = \left[ \begin{array}{ccc}
        {w_{1,2}} & {-w_{1,2}} & {0}  \\
        {-w_{1,2}} & {w_{1,2}+w_{2,3}} & {-w_{2,3}}  \\
        {0} & {-w_{2,3}} & {w_{2,3}}
        \end{array} \right],
\end{equation}
\begin{equation}
    \label{eq:Laplacian111ExampleCategory2}
    \begin{aligned}
    &\boldsymbol{L}_{n^{'}} = \\
    &\left[ \begin{array}{cccccc}
        {w_{1,2}+w_{2,3}} & {-w_{2,3}} & {-w_{1,2}} & {0} & {0} & {0}  \\
        {-w_{2,3}} & {w_{1,2}+w_{2,3}} & {0} & {-w_{1,2}} & {0} & {0}  \\
        {-w_{1,2}} & {0} & {w_{1,2}+w_{2,3}} & {0} & {-w_{2,3}} & {0}  \\
        {0} & {-w_{1,2}} & {0} & {w_{1,2}+w_{2,3}} & {0} & {-w_{2,3}}  \\
        {0} & {0} & {-w_{2,3}} & {0} & {w_{1,2}+w_{2,3}} & {-w_{1,2}}  \\
        {0} & {0} & {0} & {-w_{2,3}} & {-w_{1,2}} & {w_{1,2}+w_{2,3}}
        \end{array} \right].
    \end{aligned}
\end{equation}
The induced graphs of these partitions are depicted in figure \ref{fig:InducedGraphs4Path3}.

\section{Optimization of the Convergence Rate}
\label{sec:Lambda2SDP}

In the previous section we have modeled the Continuous Time Quantum Consensus problem as the classical CTC problem where its underlying graph is a cluster of induced graphs.
Furthermore, we have shown that the second eigenvalue of the Laplacian matrices of all the induced graphs are the same.
This eigenvalue effects the convergence rate of the CTC problem and therefore that of the Continuous Time Quantum Consensus problem.
In this section we address the optimization of the second eigenvalue of the Laplacian matrices of the induced graphs in the modeled CTC problem.
In previous section, this problem is introduced as the Fastest Continuous Time Quantum Consensus (FCTQC) problem (\ref{eq:FCTQCInitial}).
Considering the fact that the value of this eigenvalue is the same for all induced graphs then this optimization problem is reduced to optimizing the second eigenvalue the Laplacian matrix of the induced graph corresponding to partition $n=[N-1,1]$.

In the following, first we provide the optimal results for all possible connected topologies with $N=2, 3$ and $4$ vertices serving as the underlying graph of the FCTQC problem.
Then we study the FCTQC problem over quantum networks with different topologies in their general form.
We categorize these topologies into two groups.
First group, is the group of topologies that the FCTQC problem have been solved using linear programming \cite{BoydBook2004} and the second are those topologies that the FCTQC problem have been solved using semidefinite programming \cite{BoydBook2004}.
For the complete-cored star topology we have provided the detailed solution while for the rest of the topologies we only report the final optimal weights.

\subsection{The Optimal Results for all topologies with $N=2,3$ and $4$ vertices}
Here we provide the optimal weights and the second smallest eigenvalue of the Laplacian matrix for all possible topologies with $N=2,3$ and $4$ vertices which are connected and non-isomorphic.
The order of presentation of graphs here is in terms of increasing value of $\lambda_2$ for a given value of $D$.
In other words, for a given $D$ path graph has the least convergence rate while complete graph has the most.

For a network with $N=2$ vertices, the only connected topology is the path graph with $2$ vertices.
The optimal value of the second smallest eigenvalue $(\lambda_2)$ for path topology with $2$ vertices is $2D$ and the optimal weight is $D$.
In case of a network with $N=3$ vertices, there are two connected topologies, namely, path topology with $3$ vertices and the triangular topology which is a complete graph.
For the path topology with $3$ vertices, the optimal value of the second smallest eigenvalue $(\lambda_2)$ is $D/2$ and the optimal weight is $D/2$.
For the triangular topology, the optimal $\lambda_2$ and weight are $D$ and $D/3$, respectively.

\begin{figure}
  \centering
     \includegraphics[width=130mm]{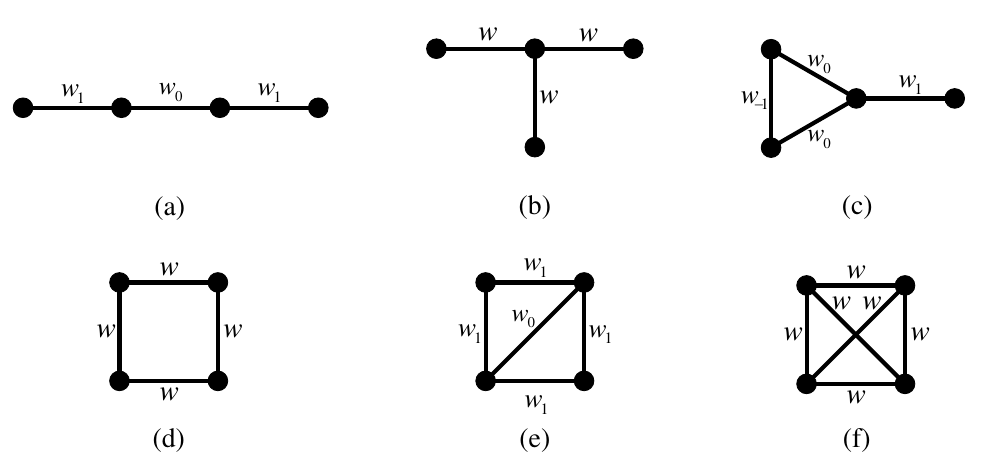}
  \caption{All possible connected underlying topologies with $N=4$ vertices which are non-isomorphic.}
  \label{fig:N4Graphs}
\end{figure}

Path graph with $N=4$ depicted in figure \ref{fig:N4Graphs} (a):
\begin{subequations}
    \label{eq:PathN4Results}
    \begin{gather}
        w_0  =  2D/5,   \quad \quad     w_1  =  3D/10,  \label{eq:PathN4ResultsW} \\
        \lambda_2  =  D/5,  \label{eq:PathN4ResultsS}
     \end{gather}
\end{subequations}
Star graph with $N=4$ depicted in figure \ref{fig:N4Graphs} (b):
\begin{subequations}
    \label{eq:StarN4Results}
    \begin{gather}
        w  =  D/3,      \label{eq:StarN4ResultsW} \\
        \lambda_2  =  D/3,  \label{eq:StarN4ResultsS}
     \end{gather}
\end{subequations}
Lollipop graph with $N=4$ depicted in figure \ref{fig:N4Graphs} (c):
\begin{subequations}
    \label{eq:LollipopN4Results}
    \begin{gather}
        w_{-1}  =  D\left( \frac{2-\sqrt{3}}{6} \right),   \quad   w_0 = D/3,   \quad   w_1 = D/2,      \label{eq:LollipopN4ResultsW} \\
        \lambda_2  =  D\left( 1-\frac{1}{\sqrt{3}} \right),  \label{eq:LollipopN4ResultsS}
     \end{gather}
\end{subequations}
Cycle graph with $N=4$ depicted in figure \ref{fig:N4Graphs} (d):
\begin{subequations}
    \label{eq:CycleN4Results}
    \begin{gather}
        w  =  D/4,      \label{eq:CycleN4ResultsW} \\
        \lambda_2  =  D/2,  \label{eq:CycleN4ResultsS}
     \end{gather}
\end{subequations}
Paw graph with $N=4$ depicted in figure \ref{fig:N4Graphs} (e):
\begin{subequations}
    \label{eq:PawN4Results}
    \begin{gather}
        w_{0}  =  0,   \quad   w_1 = D/4,         \label{eq:PawN4ResultsW} \\
        \lambda_2  =  D/2,  \label{eq:PawN4ResultsS}
     \end{gather}
\end{subequations}
Complete graph with $N=4$ depicted in figure \ref{fig:N4Graphs} (f):
\begin{subequations}
    \label{eq:CompleteN4Results}
    \begin{gather}
        w  =  D/6,      \label{eq:CompleteN4ResultsW} \\
        \lambda_2  =  2D/3,  \label{eq:CompleteN4ResultsS}
     \end{gather}
\end{subequations}
Note that for Paw graph the weight on the diameter ($w_{0}$) is zero and the optimal weights and $\lambda_2$ are same as those of Cycle graph.

\subsection{Linear Programming}
In this section, we provide the optimal results for a number of topologies where the FCTQC problem can be solved using Linear Programming \cite{BoydBook2004}.

\subsubsection{Cartesian Product of Edge Transitive Graphs}
This topology is obtained from Cartesian product of $m$ edge-transitive weighted graphs.
The weighted Laplacian matrix for the whole graph can be written as below,
\begin{equation}
    \nonumber
    \begin{gathered}
        L^{w} = \sum\nolimits_{i=1}^{m}     {   I_{N_{1}} \otimes I_{N_{2}} \otimes \cdots \otimes I_{N_{i-1}} \otimes L^{w}_{i} \otimes I_{N_{i+1}} \otimes \cdots \otimes I_{N_{m}}       }
    \end{gathered}
\end{equation}
where $L^{w}_{i}$ is the weighted Laplacian matrix of $i$-th graph.
$I_{N_{j}}$ is the identity matrix with size $N_{j}$ and $N_{j}$ is the number of vertices in the $j$-th graph.
Due to edge-transitivity, all edges of each edge-transitive graph have the same optimal weight.
Thus the weighted Laplacian matrix $(L^{w}_{i})$ for each one of the graphs can be written as $(L^{w}_{i}) = w_i \cdot L_{i}$ in terms of its unweighted Laplacian matrix $(L_{i})$.
Using this relation the weighted Laplacian matrix for the whole graph can be derived as below,
\begin{equation}
    \label{eq:CartesianLaplacian}
    \begin{gathered}
        L^{w} = \sum\nolimits_{i=1}^{m}     {   w_i \cdot I_{N_{1}} \otimes I_{N_{2}} \otimes \cdots \otimes I_{N_{i-1}} \otimes L_{i} \otimes I_{N_{i+1}} \otimes \cdots \otimes I_{N_{m}}       }
    \end{gathered}
\end{equation}
We denote the eigenvalues of the $i$-th unweighted Laplacian matrix $L_{i}$ in their sorted order by $\lambda_{i,\alpha_i}$ where $\alpha_i$ varies from $1$ to $N_i$.
Using this notation the eigenvalues of the weighted Laplacian of the whole graph can be written as below,
\begin{equation}
    \label{eq:CartesianLaplacianEigenvalue}
    \begin{gathered}
        \lambda^{w}_{\alpha_1, \alpha_1, \ldots, \alpha_m}  =  w_1 \cdot \lambda_{1,\alpha_1}  +   w_2 \cdot \lambda_{2,\alpha_2}  +  \cdots  + w_m \cdot \lambda_{m,\alpha_m}
    \end{gathered}
\end{equation}
where $\alpha_i$ for $i=1,\ldots,m$ varies from $1$ to $N_i$.
Based on the derivation in (\ref{eq:CartesianLaplacianEigenvalue}) and considering the fact that first eigenvalue of each unweighted Laplacian matrix $L_{i}$ is zero (i.e. $\lambda_{i,1} = 0$ for $i = 1, \ldots, m$), the second smallest eigenvalue of the weighted Laplacian of the whole graph can be written as $\lambda^{w}_2  =  \min { w_1 \cdot \lambda_{1,2},  w_2 \cdot \lambda_{2,2}, \cdots,   w_m \cdot \lambda_{m,2}  }$.
Using this result the optimization problem for the FCTQC problem can be written as below,
\begin{equation}
    \label{eq:CartesianLP}
    \begin{aligned}
        \max\limits_{w_1, w_2, \cdots, w_m} \quad &s=\min_{i}{w_i\cdot\lambda_{i,2}},  \\
        s.t.  \quad\quad \   &\sum_{j=1}^{m}{\tilde{E}_j \cdot w_j} = D.
    \end{aligned}
\end{equation}
where $\tilde{E}_j = E_j \cdot \prod_{\substack{k=1\\k\neq j}}^{m} {N_{k}}$ and $E_j$ is the number of edges in the $j$-th edge-transitive graph.
For the optimal answer we have
\begin{equation}
    \label{eq:CartesianLPOptimalRelaion}
    \begin{aligned}
        s = w_1 \cdot \lambda_{1,2}  =  w_2 \cdot \lambda_{2,2}  =  \cdots  =  w_m \cdot \lambda_{m,2}.
    \end{aligned}
\end{equation}
From this relation we can conclude the following for the optimal value of $\lambda_2$ and the weights
\begin{subequations}
    \label{eq:CartesianLPOptimalAnswer}
    \begin{gather}
        \lambda_2  =  s  =  \frac{D}{\tilde{N}\left( \sum\limits_{j=1}^{m}{\frac{E_j}{N_j \cdot \lambda_{j,2} }} \right)},      \label{eq:CartesianLPOptimalAnswerS} \\
        w_j  =  \frac{s}{\lambda_{j,2}} \quad \text{for} \quad j=1,\ldots,m ,  \label{eq:CartesianLPOptimalAnswerW}
     \end{gather}
\end{subequations}
where $\tilde{N} = \prod_{i=1}^{m}{N_{i}}$.

\begin{figure}
  \centering
     \includegraphics[width=80mm]{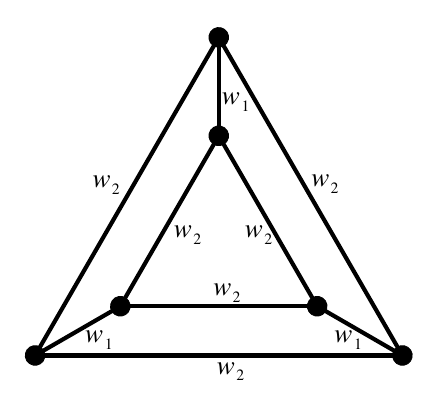}
  \caption{The Cartesian product of two complete graphs, namely, $K_2$ and $K_3$.}
  \label{fig:PrismGraph}
\end{figure}

As an example consider the Cartesian product of two complete graphs each with $N_1$ and $N_2$ vertices.
Considering $w_1$ and $w_2$ as the weights on the edges of each one of the complete graphs, then for the optimal results we have
\begin{subequations}
    \label{eq:CartesianCompleteGraphOptimalAnswer}
    \begin{gather}
        \lambda_2  =  s  =  \frac{  2D  }   {   2 N_1 N_2 - N_1 - N_2   },      \label{eq:CartesianCompleteGraphOptimalAnswerS} \\
        w_1 = s/N_1,        \quad       w_2 = s/N_2,  \label{eq:CartesianCompleteGraphOptimalAnswerW}
     \end{gather}
\end{subequations}
An obvious example for the Cartesian product of two complete graphs is the Cartesian product of $K_2$ and $K_3$ as depicted in figure \ref{fig:PrismGraph}.
This graph is also known as the Prism graph.
The optimal value of $\lambda_2$ is $2D/7$ and for the optimal weights we have $w_1 = D/7$ and $w_2 = 2D/21$.

In the following we provide the optimal answer to the FCTQC problem for two well-known edge-transitive graphs, namely the complete graph and the cycle graph.

\subsubsection{Complete Graph}
A complete graph with $N$ vertices is a graph where each node is connected to every other node in the graph.
Due to the symmetry of the graph all edges have the same weight, and its Laplacian matrix can be written as below,
\begin{equation}
    \nonumber
    \begin{gathered}
        L_W = (N-1)\cdot w \cdot \boldsymbol{I}  -  w \cdot ( \boldsymbol{J} - \boldsymbol{I} )  =  N \cdot w \cdot \boldsymbol{I} - w \cdot \boldsymbol{J}.
    \end{gathered}
\end{equation}
The eigenvalues of the Laplacian matrix for a complete graph are as following
\begin{equation}
    \nonumber
    \begin{gathered}
        \lambda_1 = 0, \lambda_{2}= \lambda_{3} = \cdots = \lambda_{N} = N \cdot w.
    \end{gathered}
\end{equation}
Considering the constraint on the summation of the weights (\ref{eq:FCTCOptmizationProblem1}) in the FCTC problem, we can conclude the following for the optimal weight on the edges of the complete graph,
\begin{equation}
    \nonumber
    \begin{gathered}
        \frac{N(N-1)}{2} w = D  \Rightarrow \lambda_2 = \frac{2D}{N-1}  .
    \end{gathered}
\end{equation}

\subsubsection{Cycle Topology}
In this topology $N$ vertices are connected in form of a cycle.
A cycle graph with four vertices in depicted in figure \ref{fig:N4Graphs} (d).
Cycle graph is edge transitive and thus the optimal weight on all edges is the same.
The optimal value of $\lambda_2$ and the weight are as below
\begin{subequations}
    \nonumber
    \begin{gather}
        w  =  D/N,      \nonumber \\ 
        \lambda_2  =  \frac     {   2(1-\cos{(2\pi/N)})D   }      {   N   },  \nonumber  \\   
     \end{gather}
\end{subequations}

\subsection{Semidefinite Programming}
In this section we provide the optimal results for a number of topologies where the FCTQC problem can be solved using semidefinite Programming \cite{BoydBook2004}.
For complete-cored symmetric star topology we provide the detailed solution while for the rest of the topologies discussed in this section we only provide the optimal results.

\subsubsection{Complete-Cored Symmetric Star Topology}

\begin{figure}
  \centering
     \includegraphics[width=130mm]{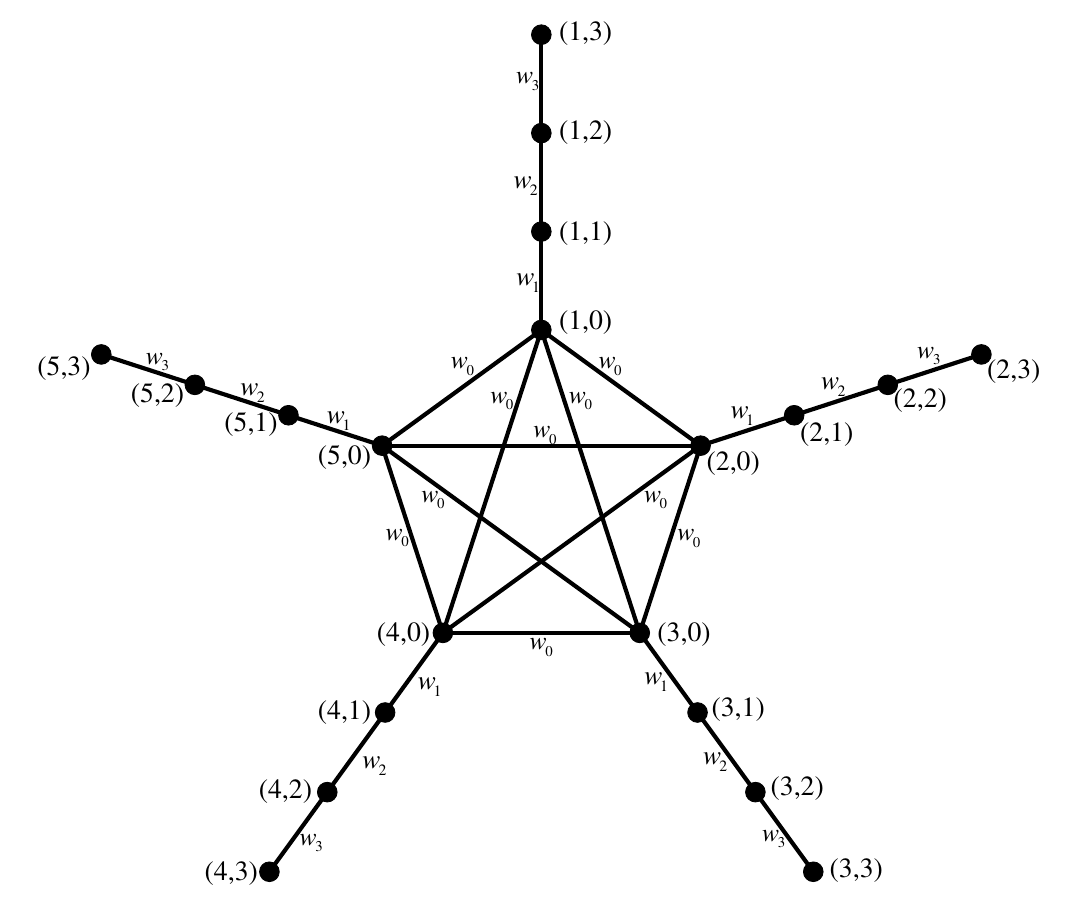}
  \caption{Weighted Complete-Cored Symmetric star topology with $p=5$ branches of length $q=3$.}
  \label{fig:CompleteCoredStar}
\end{figure}

Complete-Cored Symmetric (CCS) star topology with parameters $(p,q)$ consists of $p$ path branches of length $q$, referred to as tails.
Each one of the path branches contains $q$ edges.
Tails are connected to each other at one end to form a complete graph in the core.
A CCS star graph with parameters $p=5$, $q=3$ is depicted in figure \ref{fig:CompleteCoredStar}.

Automorphism of the CCS star graph is $S_p$ permutation of tails.
Using the stratification method \cite{SaberThesis2015} it has $q+1$ edge orbits, namely edges connecting, vertices in the complete part to each other and $q$ class of edges in tails.
Thus it suffices to consider just $q+1$ weights $w_{0}, w_{1}, \cdots, w_{q}$ (as labeled in figure \ref{fig:CompleteCoredStar}).
Defining the weight matrix accordingly and using the proper orthonormal basis (as introduced in \cite{SaberThesis2015}) the weight matrix transforms into a block diagonal matrix where the diagonal blocks in the new basis are either one of the following matrices,
\begin{equation}
    \nonumber
    \begin{gathered}
        L_0 =  \left[ \begin{array}{ccccc}
                {w_1}             &{-w_1}       &{0}          &{\cdots}         &{0}              \\
                {-w_1}            &{w_1+ w_2}   &{-w_2}       &{\cdots}         &{\vdots}         \\
                {0}               &{-w_2}       &{w_2+w_3}    &{\cdots}         &{0}              \\
                {\vdots}          &{\vdots}     &{\vdots}     &{\ddots}         &{-w_{q}}       \\
                {0}               &{\cdots}     &{0}          &{-w_{q}}         &{w_{q}}
                \end{array} \right]
    \end{gathered}
\end{equation}
\begin{equation}
    \label{eq:CCSMatrixL1Startification}
    \begin{aligned}
        L_1 =  \left[ \begin{array}{ccccc}
                {w_1+p \cdot w_0}        &{-w_1}       &{0}          &{\cdots}         &{0}              \\
                {-w_1}             &{w_1+ w_2}   &{-w_2}       &{\cdots}         &{\vdots}         \\
                {0}                &{-w_2}       &{w_2+w_3}    &{\cdots}         &{0}              \\
                {\vdots}           &{\vdots}     &{\vdots}     &{\ddots}         &{-w_q}           \\
                {0}                &{\cdots}     &{0}          &{-w_q}           &{w_q}
                \end{array} \right]
    \end{aligned}
\end{equation}
Considering the relation $L1 = L_0 + nw_0 \boldsymbol{e}_0 \times \boldsymbol{e}_0^T$ between matrices $L_0$ and $L_1$ and using the Courant-Weyl inequalities theorem \cite{CourantWeylInequality1995}, \cite{SaberThesis2015} the following corollary for the eigenvalues of $L_0$ and $L_1$ can be concluded,
\begin{equation}
    \label{eq:CCSEigenvalueRelation}
    \begin{gathered}
        \lambda_1(L_0)  \leq  \lambda_1(L_1)  \leq  \lambda_2(L_0)  \leq  \cdots \leq \lambda_{q-1}(L_0) \leq \lambda_{q-1}(L_1) \leq \lambda_{q}(L_0) \leq \lambda_{q}(L_1).
    \end{gathered}
\end{equation}
It is obvious from above relations that the second eigenvalue of the original Laplacian matrix $\lambda_2(L)$ is the smallest eigenvalue of $L_1$ (i.e. $\lambda_1$).

We define the following column vectors (each with $q+1$ elements) as basis for matrix $L_1$ (\ref{eq:CCSMatrixL1Startification})
\begin{equation}
    \nonumber
    \begin{gathered}
        \boldsymbol{e}_0 (j) =
	       \begin{cases}
                1 \quad \text{for} \quad j = 1 \\
                0 \quad \text{Otherwise}
            \end{cases} \\
        \boldsymbol{e}_i (j) =
	       \begin{cases}
                -1 \quad \text{for} \quad j = i \\
                1 \quad \text{for} \quad j = i+1 \\
                0 \quad \text{Otherwise}
            \end{cases}
            \quad\quad \text{for} \quad i=1, 2, \ldots, q.
    \end{gathered}
\end{equation}
Thus $L_1$ can be written as below,
\begin{equation}
    \nonumber
    \begin{gathered}
        L_1 = p \cdot w_0 \cdot \boldsymbol{e}_0 \times \boldsymbol{e}_{0}^{T}  +  2 \sum_{j=1}^{q}  {  w_i \cdot \boldsymbol{e}_j \times \boldsymbol{e}_j^t  }
    \end{gathered}
\end{equation}
Based on the results above we can express the Fastest Continuous Time Consensus (FCTC) problem in the form of the semidefinite programming ( as described in Appendix \ref{sec:SDP}) as below,
\begin{equation}
    \label{eq:CCSFCTCSDP}
    \begin{aligned}
        \min\limits_{w_i} \quad &-s  \\
        s.t. \quad &L_1 - s \cdot I_{q+1} \geq 0, \\
        &D - \frac{p(p-1)}{2}w_{0} - p \cdot \sum_{i=1}^{q}{w_{i}} \geq 0
    \end{aligned}
\end{equation}
In order to formulate problem (\ref{eq:CCSFCTCSDP}) in the form of standard semidefinite programming (\ref{eq:SDPPrimal}), we define $\boldsymbol{F}_i, \boldsymbol{c}$  and $\boldsymbol{x}$ as below,
\begin{equation}
    \nonumber
    \begin{gathered}
        \boldsymbol{F}_{0} =
                \left[ \begin{array}{cc}
                {\boldsymbol{0}_{(q+1)}}       &{\boldsymbol{0}}       \\
                {\boldsymbol{0}}                  &{D}
                \end{array} \right],
        \qquad
        \boldsymbol{F}_{s} =
                \left[ \begin{array}{cc}
                {-\boldsymbol{I}_{(q+1)}}       &{\boldsymbol{0}}       \\
                {\boldsymbol{0}}                &{0}
                \end{array} \right]
    \end{gathered}
\end{equation}
\begin{equation}
    \nonumber
    \begin{gathered}
        \boldsymbol{F}_{w_{0}} =
                \left[ \begin{array}{cc}
                {p \cdot \boldsymbol{e}_{0} \times \boldsymbol{e}_{0}^{T}}       &{\boldsymbol{0}}       \\
                {\boldsymbol{0}}                &{-\frac{p(p-1)}{2}}
                \end{array} \right]
    \end{gathered}
\end{equation}
\begin{equation}
    \nonumber
    \begin{gathered}
        \boldsymbol{F}_{w_{j}} =
                \left[ \begin{array}{cc}
                {2 \cdot \boldsymbol{e}_{j} \times \boldsymbol{e}_{j}^{T}}       &{\boldsymbol{0}}       \\
                {\boldsymbol{0}}                &{-p}
                \end{array} \right]
                \quad \quad \text{for} \quad \quad j = 1, \ldots, q,
    \end{gathered}
\end{equation}
\begin{equation}
    \nonumber
    \begin{gathered}
        \boldsymbol{x}  =
                \left[ \begin{array}{cccccc}
                {w_0} &{w_1} &{w_2} &{\ldots} &{w_q} &{s}
                \end{array} \right]
    \end{gathered}
\end{equation}
\begin{equation}
    \nonumber
    \begin{gathered}
        \boldsymbol{c}(j)  =
                \begin{cases}
                    0 \quad \text{for} \quad j = 1, \ldots, q+1,\\
                    -1 \quad \text{for} \quad j = q+2.
                \end{cases}
    \end{gathered}
\end{equation}
where
\begin{equation}
    \nonumber
    \begin{gathered}
        \boldsymbol{F}(\boldsymbol{x})  =   \boldsymbol{F}_{0} + s \cdot \boldsymbol{F}_{s} + w_{0} \cdot \boldsymbol{F}_{w_{0}} + \sum_{j=1}^{q}{w_{j} \cdot \boldsymbol{F}_{w_{j}}}  \geq \boldsymbol{0}
    \end{gathered}
\end{equation}
The dual problem can be written as below,
\begin{equation}
    \label{eq:CCSFCTCSDPDual}
    \begin{aligned}
        \max\limits_{\boldsymbol{Z}} \quad &-Tr[\boldsymbol{F}_0 \times \boldsymbol{Z}],  \\
        s.t. \quad &\boldsymbol{Z} \succeq \boldsymbol{0}, \\
        &Tr[ \boldsymbol{F}_{s} \times \boldsymbol{Z} ] = \boldsymbol{c}_{q+2} = -1 \\
        &Tr[\boldsymbol{F}_{w_i} \times \boldsymbol{Z}] = \boldsymbol{c}_{i+1} = 0 \quad \text{for} \quad i=0, 1, \ldots, q.
    \end{aligned}
\end{equation}
Since the set of basis $\{e_\mu|\mu=0,1,\ldots,q\}$ are linearly independent one can introduce their dual as $\{\widetilde{\boldsymbol{e}}_\mu|\mu=0,1,\ldots,q\}$ such that $ {\widetilde{\boldsymbol{e}}_\mu}^T \times e_\nu = e_{\nu}^{T} \widetilde{\boldsymbol{e}}_\mu=\delta_{\mu,\nu}$.
The dual variable $\boldsymbol{Z}$ can be written in terms of the dual basis as below,
\begin{equation}
    \label{eq:CCSFCTCSDPZExpansion}
    \begin{aligned}
        \boldsymbol{Z}  =
                \left[ \begin{array}{c}
                { \sum_{\mu=0}^{q}{z_{\mu} \times \widetilde{\boldsymbol{e}}_\mu }  } \\
                { z_{D} }
                \end{array} \right]
                \times
                \left[ \begin{array}{cc}
                { \sum_{\mu=0}^{q}{z_{\mu} \times \widetilde{\boldsymbol{e}}_\mu^{T} }  } &  { \hspace{10pt} z_{D} }
                \end{array} \right]
    \end{aligned}
\end{equation}
Using the expansion (\ref{eq:CCSFCTCSDPZExpansion}), the dual constraints reduce to the following
\begin{subequations}
    \label{eq:CCSFCTCSDPDualConstraints}
    \begin{gather}
        Tr[\boldsymbol{F}_{w_0} \times \boldsymbol{Z}]  = 0\quad \Rightarrow \quad p|z_0|^2-\frac{p(p-1)}{2}|z_D|^2 =0,  \label{eq:CCSFCTCSDPDualConstraintsA} \\
        Tr[\boldsymbol{F}_{w_i} \times \boldsymbol{Z}]  = 0 \quad \Rightarrow  \quad 2|z_j|^2-p|z_D|^2 =0,\quad \text{for} \quad j=1,2,\cdots,q,  \label{eq:CCSFCTCSDPDualConstraintsB}
     \end{gather}
\end{subequations}
where we can conclude that,
\begin{equation}
    \label{eq:CCSFCTCSDPDualzRelations}
    \begin{gathered}
        p \cdot |z_0|^2= (p-1) \cdot |z_j|^2, \quad \text{for} \quad j=1,2,\cdots,q
    \end{gathered}
\end{equation}
Using dual relations (\ref{eq:CCSFCTCSDPDualzRelations}) the complementary slackness condition (\ref{eq:SDPOptimality}) reduces to
\begin{subequations}
    \label{eq:CCSFCTCSDPSlacknessRelations}
    \begin{align}
        p w_0 z_0 \boldsymbol{e}_0 + 2\sum_{j=1}^q {w_j z_j \boldsymbol{e}_j}  =  s(\sum_{\mu=0}^q {z_\mu\widetilde{\boldsymbol{e}}_\mu}),  \label{eq:CCSFCTCSDPSlacknessRelationsA} \\
        z_D \left( D - \frac{p(p-1)}{2} w_0 - p\sum_{j=1}^q {w_j} \right) = 0  \label{eq:CCSFCTCSDPSlacknessRelationsB}
     \end{align}
\end{subequations}
Setting $z_D = 0$ is not acceptable since it will lead to the trivial case of $z_0 = z_1 = \cdots = z_q = 0$, thus equation (\ref{eq:CCSFCTCSDPSlacknessRelationsB}) reduces to
\begin{equation}
    \label{eq:CCSFCTCSDPSlacknessRelationsZD}
    \begin{gathered}
        D - \frac{p(p-1)}{2} w_0 - p\sum_{j=1}^{q} {w_j} = 0
    \end{gathered}
\end{equation}
By multiplying both sides of (\ref{eq:CCSFCTCSDPSlacknessRelationsA}) by vectors $e_\nu$ for $\nu=0,1,\cdots,q$ we have
\begin{equation}
    \label{eq:CCSFCTCSDPSlacknessRelationsGramForm}
    \begin{gathered}
        p \boldsymbol{G}_{\nu,0} w_0 z_0  +  2 \sum_{j=1}^{q} {\boldsymbol{G}_{\nu,j} w_j z_j = s z_\nu },\quad \text{for} \quad \nu=0,1,2,\ldots,q,
    \end{gathered}
\end{equation}
where $\boldsymbol{G}$ is the gram matrix and it is defined as $\boldsymbol{G}_{\mu,\nu} = \boldsymbol{e}_{\mu}^{T} \times \boldsymbol{e}_{\nu} = \boldsymbol{e}_{\nu}^{T} \times \boldsymbol{e}_\mu = \boldsymbol{G}_{\nu,\mu}$.
The gram matrix can be written as below,
\begin{equation}
    \nonumber
    \begin{gathered}
        \boldsymbol{G}  =
                \left[ \begin{array}{cccccc}
                {1}                     &{-\frac{1}{\sqrt(2)}}  &{0}            &{\cdots}       &{0}  \\
                {-\frac{1}{\sqrt(2)}}   &{1}                    &{-\frac{1}{2}} &{\cdots}       &{\vdots}  \\
                {0}                     &{-\frac{1}{2}}         &{1}            &{\cdots}       &{0}  \\
                {\vdots}                &{\vdots}               &{\vdots}       &{\ddots}       &{-\frac{1}{2}}  \\
                {0}                     &{\cdots}               &{0}            &{-\frac{1}{2}} &{1}
                \end{array} \right].
    \end{gathered}
\end{equation}
Based on (\ref{eq:CCSFCTCSDPDualzRelations}), we can assume that $z_0 = \sqrt{\frac{n-1}{2}} \cdot z_{D}$, $z_j = \sqrt{\frac{n}{2}} \cdot z_{D}$ for $j=1,2,\cdots q$.
Substituting these values in (\ref{eq:CCSFCTCSDPDualConstraints}) we obtain the following for the optimal weights,
\begin{subequations}
    \label{eq:CCSFCTCOptimalWeights}
    \begin{gather}
        w_0 = s\frac { (p-1)\left(\boldsymbol{G}^{-1}\right)_{0,0} + \sqrt{p(p-1)}\sum_{i=1}^{q} {\left(\boldsymbol{G}^{-1}\right)_{0,i}} }  {p(p-1)},  \label{eq:eq:CCSFCTCOptimalWeightsW0} \\
        w_j  =  s \frac {\sqrt{p(p-1)} \left(\boldsymbol{G}^{-1}\right)_{j,0} + p \sum_{i=1}^{q} {\boldsymbol{G}^{-1}_{j,i}} } {2p} \quad \text{for} \quad j=1,2,\cdots,q.  \label{eq:CCSFCTCOptimalWeightsWi}
     \end{gather}
\end{subequations}
Substituting the obtained optimal weights (\ref{eq:CCSFCTCOptimalWeights}) in equation (\ref{eq:CCSFCTCSDPSlacknessRelationsZD}) we obtain the following equation for the optimal value of the second smallest eigenvalues $\lambda_2$,
\begin{equation}
    \label{eq:CCSFCTCOptimalLambda2First}
    \begin{aligned}
        &\lambda_2 = s =
        \\
        &\frac{2D}{(p-1)\left(\boldsymbol{G}^{-1}\right)_{0,0} + 2\sqrt{p(p-1)}\sum_{i=1}^q {\left(\boldsymbol{G}^{-1}\right)_{0,i}} + p\sum_{i,j=1}^{q} {\left(\boldsymbol{G}^{-1}\right)_{j,i}}}
    \end{aligned}
\end{equation}
where $\boldsymbol{G}^{-1}$ is inverse of Gram matrix.
Note that matrices $\boldsymbol{G}$ and $\boldsymbol{G}^{-1}$ are square matrices of dimension $q+1$ but we have started the index of elements from $0$ (e.g. $\left(\boldsymbol{G}^{-1}\right)_{0,0}$).
This notation has been used due to the fact that the element in the first row and first column of the Gram matrix $\boldsymbol{G}$ refers to the inner product of the vector $\boldsymbol{e}_{0}^{T}$ and $\boldsymbol{e}_{0}$.
Thus the notation $\left(\boldsymbol{G}^{-1}\right)_{0,0}$ refers to the elements on first row and first column of the inverse Gram matrix $(\boldsymbol{G}^{-1})$.
The inverse Gram matrix $(\boldsymbol{G}^{-1})$ is as below,
\begin{equation}
    \label{eq:CCSFCTCOptimalInverseGram}
    \begin{gathered}
            (G^{-1})_{i,j}  =
            \begin{cases}
                q+1 \quad \text{for} \quad j = i = 0 \\
                \sqrt{2}(q-j+1) \quad \text{for} \quad i=0, j=1, \ldots, q  \\
                \sqrt{2}(q-i+1) \quad \text{for} \quad j=0, i=1, \ldots, q  \\
                2\max(q-i+1,q-j+1) \quad \text{for} \quad i,j = 1, \ldots, q
            \end{cases}
    \end{gathered}
\end{equation}
Substituting the inverse Gram matrix  in (\ref{eq:CCSFCTCOptimalLambda2First}) we obtain the following results for the optimal value of the second smallest eigenvalue $\lambda_{2}$ and the optimal weights,
\begin{subequations}
    \label{eq:CCSFCTCOptimalLambda2Second}
    \begin{align}
        &w_0  =  \frac 	{	3D  \left( 2p - 2 + q\sqrt{2p(p-1)} \right) 	} 	{	p(p-1) \left( 3p - 3 + 3q\sqrt{2p(p-1)}  +  2pq^2  +  pq \right)	} \label{eq:CCSFCTCOptimalLambda2SecondW0} \\
        &w_j  =  \frac 	{   3D  \left(  \sqrt{2p(p-1)} ( q-j+1 ) + p(q-j+1)(q+j)  \right)	}     {	3p(q+1)\left( p-1 + q\sqrt{2p(p-1)} \right)  +  p^{2}q(q+1)(2q+1)	}     \label{eq:CCSFCTCOptimalLambda2SecondWj} \\
        &\qquad\qquad\qquad\qquad\qquad\quad\text{for} \quad j = 1, \ldots, q,        \nonumber   \\
        &\lambda_2 = \frac {6D} {3(p-1)(q+1)+3\sqrt{2p(p-1)}q(q+1)+pq(q+1)(2q+1)}.  \label{eq:CCSFCTCOptimalLambda2SecondLambda2}
    \end{align}
\end{subequations}
A special case of the CCS star topology is the path topology with even number of vertices which is obtained for $p=2$.
The optimal weights and the optimal value of the second smallest eigenvalue $\lambda_{2}$ for the path topology with $2(q+1)$ vertices are as below,
\begin{subequations}
    \label{eq:PathEvenFCTCOptimalLambda2Second}
    \begin{align}
        &w_0  =  \frac 	{	3D  (q+1) 	} 	{	(2q+3)(2q+1)	} \label{eq:PathEvenFCTCOptimalLambda2SecondW0} \\
        &w_j  =  \frac 	{   3D  \left(  (q+1)^2 - j^2  \right)	}     {	(q+1)(2q+1)(2q+3)	}   \quad \text{for} \quad j = 1, \ldots, q,        \label{eq:PathEvenFCTCOptimalLambda2SecondWj} \\
        &\lambda_2 = \frac {6D} {(q+1)(2q+1)(2q+3)}.  \label{eq:PathEvenFCTCOptimalLambda2SecondLambda2}
    \end{align}
\end{subequations}
This result is in agreement with that of Fiedler in \cite{Fiedler1990} for a path with even number of vertices, (i.e by substituting $p=2$ and $D=2q+1$ in \cite{Fiedler1990}), which results in
\begin{subequations}
    \label{eq:CCSFCTCOptimalLambda2SecondFiedler}
    \begin{align}
        &w_0  =  \frac 	{	3  (q+1) 	} 	{	2q+3	} \label{eq:CCSFCTCOptimalLambda2SecondFiedlerW0} \\
        &w_j  =  \frac 	{   3  \left(  (q+1)^2 - j^2  \right)	}     {	(q+1)(2q+3)	}   \quad \text{for} \quad j = 1, \ldots, q,  \label{eq:CCSFCTCOptimalLambda2SecondFiedlerWj} \\
        &\lambda_2 = \frac{6}{(q+1)(2q+3)}.  \label{eq:CCSFCTCOptimalLambda2SecondFiedlerLambda2}
    \end{align}
\end{subequations}

\subsubsection{CCS Star with two types of branches}

\begin{figure}
  \centering
     \includegraphics[width=130mm]{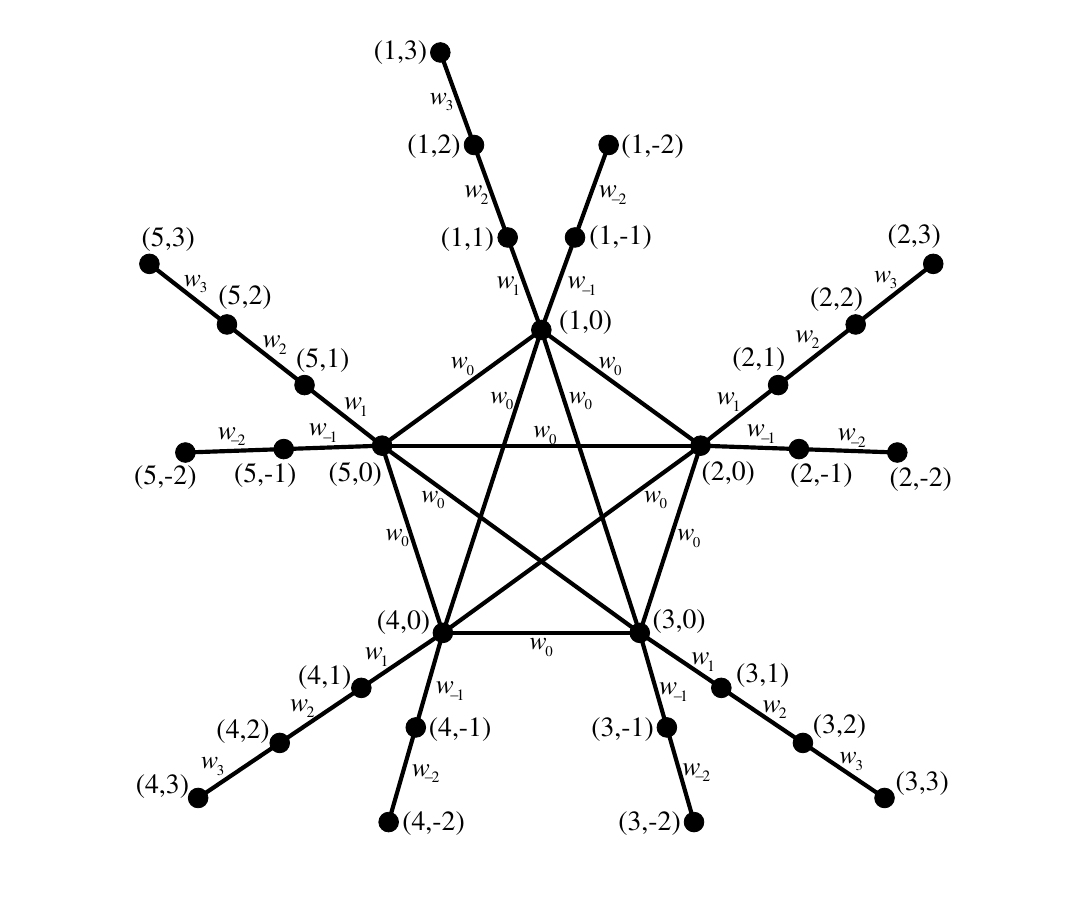}
  \caption{The weighted graph of CCS star graph with two types of branches with $p=5$ branches of length $q_1=2$ and $q_2=3$.}
  \label{fig:CompleteCoredStar2Tails}
\end{figure}

The Complete-Cored Symmetric (CCS) star with two types of branches is identified with parameters $(p, q_1, q_2)$.
This topology is a CCS star topology where two types of tails (each with $q_1$ and $q_2$ edges) are connected to each node in the complete core.
A CCS star graph  with two types of branches with parameters $p=5$, $q_1=2$ and $q_2 = 3$ is depicted in figure \ref{fig:CompleteCoredStar2Tails}.
The optimal weights for this topology are as below,
\begin{subequations}
    \label{eq:CCS2BranchFCTCOptimalWeights}
    \begin{align}
        &w_0 = s \times \frac { (p-1)\left(\boldsymbol{G}^{-1}\right)_{0,0} + \sqrt{p(p-1)}\sum\limits_{\substack{i=-q_1 \\ i \neq 0}}^{q_2} {\left(\boldsymbol{G}^{-1}\right)_{0,i}} }  {p(p-1)},  \label{eq:eq:CCS2BranchFCTCOptimalWeightsW0} \\
        &w_j  =  s \times \frac {\sqrt{p(p-1)} \left(\boldsymbol{G}^{-1}\right)_{j,0} + p \sum\limits_{\substack{i=-q_1 \\ i \neq 0}}^{q_2} {\boldsymbol{G}^{-1}_{j,i}} } {2p}  \label{eq:CCS2BranchFCTCOptimalWeightsWi}   \\
        &\qquad\qquad \text{for} \quad j=-q_1,\cdots,q_2, \quad j \neq 0,  \nonumber
     \end{align}
\end{subequations}
and for the optimal value of the second smallest eigenvalue $(\lambda_2)$ we have
\begin{equation}
    \label{eq:CCS2BranchFCTCOptimalLambda2First}
    \begin{aligned}
        &\lambda_2 = s = \\
        &\frac{2D}{(p-1)\left(\boldsymbol{G}^{-1}\right)_{0,0} + 2\sqrt{p(p-1)}\sum\limits_{\substack{i=-q_1 \\ i \neq 0}}^{q^2} {\left(\boldsymbol{G}^{-1}\right)_{0,i}} + p\sum\limits_{\substack{j,i=-q_1 \\ j,i \neq 0}}^{q_2} {\left(\boldsymbol{G}^{-1}\right)_{j,i}}}.
    \end{aligned}
\end{equation}
Both the Gram matrix $(\boldsymbol{G})$ and its inverse $(\boldsymbol{G}^{-1})$ are square matrices of dimension $q_1+q_2+1$.
As shown in figure \ref{fig:CompleteCoredStar2Tails}, the index of the weights on first type of branches (with $q_1$ edges) starts from $w_{-q_1}$.
Therefore in our notation here we have used negative indexes to refer to the weights on these edges.
2In case of the Gram matrix and its reverse the index $(-q_1, -q_1)$ refers to the element on first row and first column while the index $(0,0)$ refers to the element on the $(q_1+1)$-th row and $(q_1+1)$-th column of the matrix.
The Gram matrix $(\boldsymbol{G})$ for the CCS Star topology with two types of branches is as below
\begin{equation}
    \nonumber
    \begin{gathered}
        \boldsymbol{G}_{i,j}  =
        \begin{cases}
            -1/2 \quad \text{for} \quad j = i+1, i=-q_{1}, \ldots, q_{2}-1, \quad i \neq -1, 0 \\
            -1/2 \quad \text{for} \quad j = i-1, i=-q_{1}+1, \ldots, q_{2}, \quad i \neq 0, 1 \\
            -1/\sqrt{2} \quad \text{for} \quad \{i,j\} = \{-1,0\}, \{0,1\}, \{0,-1\}, \{1,0\} \\
            1 \quad \text{for} \quad i = j, i=-q_{1}, \ldots, q_{2}, \\
            0 \quad\quad\quad \text{Otherwise}
        \end{cases}
    \end{gathered}
\end{equation}
Substituting the Gram matrix and its inverse in (\ref{eq:CCS2BranchFCTCOptimalLambda2First}) we obtain the following formula for the second smallest eigenvalue $(\lambda_2)$ of the Laplacian matrix,
\begin{equation}
    \label{eq:CCS2BranchFCTCOptimalLambda2Second}
    \begin{gathered}
        \lambda_2 = s = \frac{6D}  {    3(p-1)(q_1 + q_2 + 1) + 3\sqrt{2p(p-1)}D_1  +  p D_2    } \\
    \end{gathered}
\end{equation}
with $D_1$ and $D_2$ as below,
\begin{equation}
    \nonumber
    \begin{gathered}
        D_1  =   q_1(q_1+1) + q_2(q_2+1),   \\
        D_2  =   q_1 ( q_1 + 1 ) ( 2q_1 + 1 )  +   q_2 ( q_2 + 1 ) ( 2q_2 + 1 ).
    \end{gathered}
\end{equation}

\subsubsection{Symmetric Star Topology}
\begin{figure}
  \centering
     \includegraphics[width=80mm]{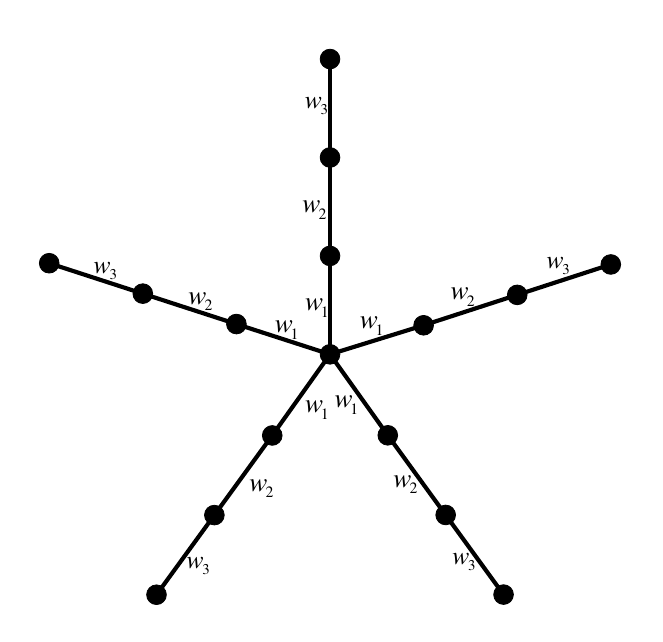}
  \caption{A symmetric star graph with $p=5$ branches of length $q=3$.}
  \label{fig:SymmetricStarGraph}
\end{figure}

A symmetric star graph with parameters $p$ and $q$ consists of $p$ path graphs (each with $q$ vertices) connected to one central vertex.
This graph has $1+pq$ vertices and $pq$ edges.
A symmetric star graph with parameters $p=5$ and $q=3$ is depicted in figure \ref{fig:SymmetricStarGraph}.
Due to the symmetry of the graph all edges with the same distance from the central vertex have the same weight, denoted by $w_j$ for $j=1, \ldots, q$.
The optimal value of the second smallest eigenvalue $(\lambda_2)$ of Laplacian matrix is
\begin{equation}
    \nonumber
    \begin{gathered}
        \lambda_2  =  \frac{6D}{pq(q+1)(2q+1)},
    \end{gathered}
\end{equation}
and the optimal value of the weights are as below,
\begin{equation}
    \nonumber
    \begin{gathered}
        w_j  =  \frac{3D(q+j)(q-j+1)}{pq(q+1)(2q+1)}, \quad \text{for} \quad j=1, \ldots, q.
    \end{gathered}
\end{equation}

\subsubsection{Palm Topology}
\label{sec:PalmGraph}
\begin{figure}
  \centering
     \includegraphics[width=80mm]{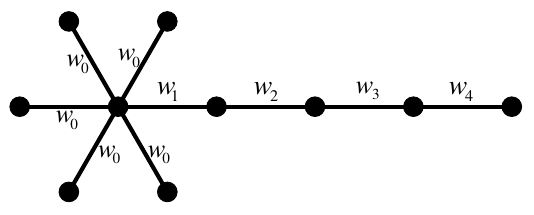}
  \caption{A palm graph with parameters $p=5$ and $q=4$.}
  \label{fig:PalmGraph}
\end{figure}

A palm graph with parameters $(p,q)$ consists of a path graph with $q$ vertices connected to the central vertex of a star graph with $p$ branches as shown in figure \ref{fig:PalmGraph} for parameters $p=5$ and $q=4$.
This graph has $p+q+1$ vertices and $p+q$ edges.
Due to the symmetry of the graph all edges connected to the central vertex of the star graph have the same weight (denoted by $w_0$), except the edge connecting the path graph to the central vertex.
The optimal answer varies, depending on the values of the parameters $p$ and $q$.
For $2p > q(q+1)$ the optimal value of the second smallest eigenvalue $(\lambda_2)$ is as below,
\begin{equation}
    \nonumber
    \begin{gathered}
        \lambda_2  =  s  =  \frac { 6D }  {  6p + q(q+1)(2q+1)  }  ,
    \end{gathered}
\end{equation}
and the optimal weights are as following,
\begin{subequations}
    \nonumber
    \begin{gather}
        w_0  =  s,  \nonumber \\
        w_j  =  \frac  {   (q-j+1) \left( (q+1)(2q+1)+p(q+j) \right)    }   {   2(p+q+1)    },  \quad \text{for} \quad j=1,\cdots,q.  \nonumber
     \end{gather}
\end{subequations}
For $2p \leq q(q+1)$ the optimal value of the second smallest eigenvalue $(\lambda_2)$ is as below,
\begin{equation}
    \nonumber
    \begin{gathered}
        \lambda_2  =  s  =  \frac { 12D(p+q+1) }  {  (q+1)(q+2)\left(6 + q(q+4p+1)\right)  }  ,
    \end{gathered}
\end{equation}
and the optimal weights are as following,
\begin{subequations}
    \nonumber
    \begin{gather}
        w_0  =  s   \frac { (q+1)(q+2) }    { 2(p+q+1) } ,  \nonumber \\
        w_j  =  s   \frac { ( q-j+1 ) \left( p(q+j+2) +(q+1)j \right) }    { 2(p+q+1) },    \quad \text{for} \quad j=1,\cdots,q. \nonumber
     \end{gather}
\end{subequations}

\subsubsection{Lollipop Topology}
\begin{figure}
  \centering
     \includegraphics[width=90mm]{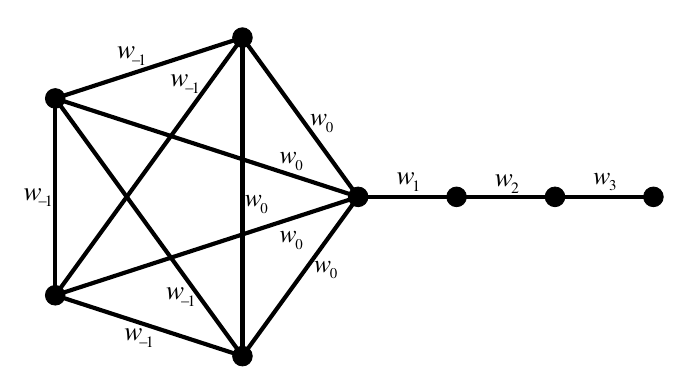}
  \caption{The weighted Lollipop graph with parameters $p=4$ and $q=3$.}
  \label{fig:CompleteGraphWithTail}
\end{figure}

This topology is obtained by connecting a path graph (with $q$ vertices) to one of the vertices in a complete graph with $p+1$ vertices.
By bridging vertex, we refer to the vertex in complete graph that is connected to the path graph.
Considering the symmetry of the complete graph the edges in the complete graph can be categorized into two groups.
The first group is those connecting the vertices in the complete graph other than the bridging vertex.
We denote the weight on the edges of the first group by $w_{-1}$.
The second group is the edges connecting the bridging vertex to other vertices in the complete graph.
We denote the weight on the edges of the second group by $w_{0}$.
The weights on the edges of the tail are denoted by $w_{1}, w_{2}, \ldots, w_{q}$.
The Lollipop topology is depicted in figure \ref{fig:CompleteGraphWithTail} along with the weights assigned to the edges.
The optimal answer varies, depending on the values of the parameters $p$ and $q$.
For $q(q+1) \leq \sqrt{2p(p+1)}$ the optimal value of the second smallest eigenvalue $(\lambda_2)$ is as below,
\begin{equation}
    \nonumber
    \begin{gathered}
        \lambda_2 = s = \frac{12D(p+q+1)}{A}.
    \end{gathered}
\end{equation}
where $A$ is
\begin{equation}
    \nonumber
    \begin{aligned}
        &A = 6(p-1)(p+q+1) \\
        &\qquad+ (q+1) \left( 6q \sqrt{2p(p+1)} + 6(p+1) + pq(2q+1) + q(q^2-1)  \right),
    \end{aligned}
\end{equation}
and the optimal weights are as following,
\begin{subequations}
    \nonumber
    \begin{gather}
        w_0  =  s \frac{q+1}{2(p+q+1)(p+1)} \left( 2(p+1) + q\sqrt{2p(p+1)} \right),  \nonumber \\
        w_{-1}  =  \frac{s-w_0}{p},  \nonumber \\
        w_j  =  s  \frac  {   (q-j+1)     }   {   2(p+q+1)    } \left(  \sqrt{2p(p+1)}  +  p(q+j) +q+1  \right),  \quad \text{for} \quad j=1,\cdots,q.  \nonumber
     \end{gather}
\end{subequations}
For the case that $q(q+1) > \sqrt{2p(p+1)}$ the optimal value of $w_{-1}$ is zero and the Lollipop topology reduces to Palm topology where the optimal answer for this topology is provided in section \ref{sec:PalmGraph}.

\subsubsection{Two Coupled Complete Graphs}
\begin{figure}
  \centering
     \includegraphics[width=90mm]{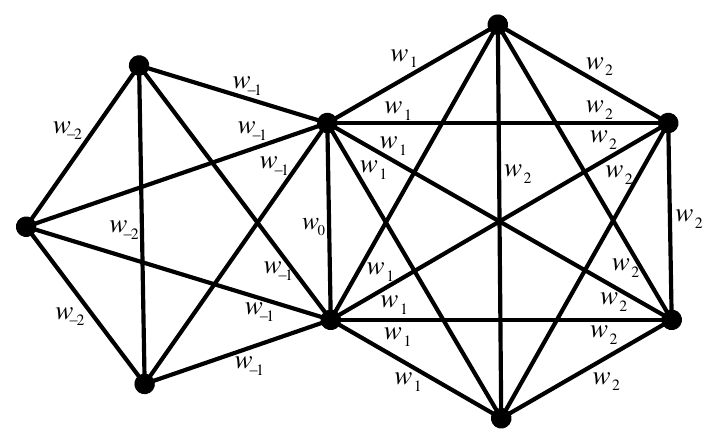}
  \caption{The weighted Lollipop graph with parameters $p=4$ and $q=3$.}
  \label{fig:TwoCoupledCompleteGraphs}
\end{figure}

In this topology, two complete graphs each with $N_1 + N_2$ and $N_2 + N_3$ vertices respectively, share $N_2$ vertices.
In figure \ref{fig:TwoCoupledCompleteGraphs} two coupled complete graphs with parameters $N_1 = 3$, $N_2 = 2$ and $N_3 = 4$ is depicted.
Due to the symmetry of the complete graphs weights can be divided into five groups.
$w_{-2}$ is the weight on edges connecting the $N_1$ vertices on the left complete graph to each other and $w_{-1}$ is the weight on the edges connecting the $N_1$ vertices on the left complete graph to the $N_2$ vertices in the middle.
$w_0$ is the weight on edges connecting the $N_2$ vertices in the middle to each other.
Similarly the weights $w_1$ and $w_2$ are defined for the weights on the edges of the complete graph on the right-hand side of the topology.
For the symmetric case where $N_1 = N_3$ the optimal weights and $\lambda_2$ are obtained for two following categories.
If $N_1 < N_2/2$ the optimal weights and $\lambda_2$ are as below
\begin{subequations}
    \nonumber
    \begin{gather}
        \lambda_2 = s = \frac       {   2 N_2 D   }           {   4 N_1 N_2 + (N_2 - 1)(N_2 - 2N_1)   },        \nonumber   \\
        w_2 = w_{-2} = 0       \qquad
        w_1 = w_{-1} = s / N_{2},             \qquad
        w_0 = ( N_2 - 2N_1 ) / N_{2}^{2},             \nonumber
     \end{gather}
\end{subequations}
If $N_1 \geq N_2/2$ the optimal weights and $\lambda_2$ are as below
\begin{subequations}
    \nonumber
    \begin{gather}
        \lambda_2 = D / (2 N_{1}),        \nonumber   \\
        w_2 = w_{-2} = 0       \qquad
        w_1 = w_{-1} = D / (2N_{1}N_{2}),             \qquad
        w_0 = 0,             \nonumber
     \end{gather}
\end{subequations}
From the optimal weights, it is apparent that for the last symmetric case where $N_1 \geq N_2/2$ the whole topology reduces to a 3-partite graph.
For the nonsymmetric case where $N_1 \neq N_3$, the optimal results are too long to report here.
But interestingly in the nonsymmetric case if $N_1 > N_3$ then the optimal value of the weight $w_{2}$ is zero.

\section{Conclusions}

We have optimized the continuous time quantum consensus algorithm in terms of its convergence rate over a quantum network with $N$ qudits.
It is shown that the optimal convergence rate is independent of the value of $d$ in qudits.
By establishing the intertwining relation between one level dominant partitions in the Hasse Diagram of integer $N$, we have shown that the spectrum of the induced graph corresponding to the dominant partition is included in that of the less dominant partition.
Based on this result, the proof of the Aldous' conjecture is extended to all possible induced graphs and it is shown that the problem of optimizing the convergence rate of the quantum consensus reduces to optimizing the second smallest eigenvalue of the Laplacian of the induced graph corresponding to partition $(N-1,1)$.
By analytically addressing the semidefinite programming formulation of the reduced optimization problem, closed-form expressions for the optimal convergence rate and the optimal weights are provided.
Interestingly, the optimal weight over some edges in certain topologies, namely Lollipop, Paw and two coupled complete Graphs are zero. This suggests that adding new edges does not necessarily improve the convergence rate.

By extending the Aldous' conjecture to all possible induced graphs, we have shown that the values of their second smallest eigenvalues are all equal. This shows that despite being disconnected and having different topologies, the induced graphs have the same spectral gap and therefore, the same asymptotic convergence rate.

\begin{appendices}

\section{Generalized Gell-Mann Matrices}
    \label{sec:GellMannMatrices}

First we introduce the $d$-dimensional elementary matrices $\{ e_{j}^{k} | k,j = 1, \ldots, d \}$.
These matrices are $d \times d$ square matrices where the element in $k$-th row and $j$-th column is one, and other elements are zero.
These matrices satisfy the commutation relation, i.e.
\begin{equation}
    \nonumber
    \begin{gathered}
        \left[ e_{j}^{i}, e_{l}^{k} \right]  =  \delta_{k,j}e_{l}^{i}  -  \delta_{i,l}e_{j}^{k}.
     \end{gathered}
\end{equation}
Based on the elementary matrices $(e_{j}^{k})$ we define the following matrices
\begin{subequations}
    \label{eq:DecompositionTetaBeta}
    \begin{align}
        \Theta_{j}^{k} &= e_{j}^{k} + e_{k}^{j},  \label{eq:DecompositionTeta} \\
        \beta_{j}^{k} &= -i\left( e_{j}^{k} - e_{k}^{j} \right)  \label{eq:DecompositionBeta} \\
        1 &\leq k < j\leq d  \nonumber
     \end{align}
\end{subequations}

In addition to off-diagonal generators, we define $d-1$ diagonal generator matrices as below
\begin{equation}
    \label{eq:DecompositionEta}
    \begin{gathered}
        \eta_{r}^{r} =  \sqrt{ \frac{2}{r(r+1)} }  \left[ \sum_{j=1}^{r} {e_{j}^{j}} - r e_{r+1}^{r+1} \right],
     \end{gathered}
\end{equation}
The range of $r$ varies from $1$ to $d-1$.
In total $d^2 - 1$ generators are defined in (\ref{eq:DecompositionTetaBeta}) and (\ref{eq:DecompositionEta}). 

Now we define the generalized Gell-Mann matrices (also known as $\lambda$ matrices) as below,
\begin{subequations}
    \label{eq:DecompositionLambdaMatrices}
    \begin{gather}
    \lambda_{(j-1)^2 + 2(k-1)}  =  \Theta_{j}^{k},   \label{eq:DecompositionLambdaMatricesa}  \\
    \lambda_{(j-1)^2 + 2k-1}  =  \beta_{j}^{k},   \label{eq:DecompositionLambdaMatricesb} \\
    \lambda_{j^2 - 1}  =  \eta_{j-1}^{j-1},   \label{eq:DecompositionLambdaMatricesc}
    \end{gather}
\end{subequations}
Note that the following relation exist between the traces of the lambda matrices defined above,
\begin{equation}
    \label{eq:DecompositionTraces}
    \begin{gathered}
        tr\{ \lambda_{\mu} \times \lambda_{\upsilon} \}  =  2 \delta_{\mu,\upsilon} \quad \text{for} \quad \mu, \upsilon \in \{0, 1, \ldots, d^2 - 1\}.
     \end{gathered}
\end{equation}
In order to form a complete orthogonal Hermitian operator basis, we add $\lambda_{0}$ defined as below to the set of $\lambda$ matrices defined in (\ref{eq:DecompositionLambdaMatrices}),
\begin{equation}
    \label{eq:DecompositionLambda0}
    \begin{gathered}
        \lambda_{0} = \sqrt{\frac{2}{d}} I_{d}.
     \end{gathered}
\end{equation}

The swapping operator $(U_{j,k})$ in terms of generalized Gell-mann matrices can be written as below,
\begin{equation}
    \label{eq:PermutationGellMann}
    \begin{aligned}
        &U_{j,k} = \\
        &\frac{1}{2}   \sum_{\mu=0}^{d^2 - 1}  { \left( I_{d} \otimes I_{d} \otimes \cdots \otimes \underbrace{\scriptstyle\lambda_{\mu}}_{j\text{-th}} \otimes I_{d} \otimes \cdots \otimes I_{d} \otimes \underbrace{\scriptstyle\lambda_{\mu}}_{k\text{-th}} \otimes I_{d} \otimes \cdots \otimes I_{d} \right)  }      +  \frac{I_{d}^{N}}{d}
     \end{aligned}
\end{equation}

As an example $SU(2)$ generators (known as Pauli matrices) are given as below,
\begin{equation}
    \nonumber
    \begin{gathered}
        \lambda_0 = I_{2}  =  \left[ \begin{array}{cc}  1 & 0  \\  0 & 1  \end{array} \right], \\
        \lambda_1 = \Theta_{2}^{1} = e_{2}^{1} + e_{1}^{2} = \left[ \begin{array}{cc}  0 & 1  \\  1 & 0  \end{array} \right], \\
        \lambda_2 = \beta_{2}^{1} = -i(e_{2}^{1} - e_{1}^{2}) = \left[ \begin{array}{cc}  0 & -i  \\  i & 0  \end{array} \right], \\
        \lambda_3 = \eta_{1}^{1} = e_{1}^{1} - e_{2}^{2} = \left[ \begin{array}{cc}  1 & 0  \\  0 & -1  \end{array} \right].
     \end{gathered}
\end{equation}

\section{Semidefinite Programming (SDP)}
    \label{sec:SDP}

Semidefinite Programming is a convex optimization problem that aims to minimize a linear function subject to a linear matrix inequality constraint \cite{BoydBook2004}.
It can be formulated as below,
\begin{equation}
    \label{eq:SDPPrimal}
    \begin{aligned}
        \min\limits_{\boldsymbol{x}} \quad &\boldsymbol{c}^{T} \cdot \boldsymbol{x} \\
        s.t. \quad &\boldsymbol{F}(\boldsymbol{x}) = \sum_{i=1}^{|\boldsymbol{x}|} { \boldsymbol{x}_{i} \boldsymbol{F}_{i} + \boldsymbol{F}_{0} }   \succeq  0,
    \end{aligned}
\end{equation}
The formulation above is referred to as the primal problem.
The minimization variable is the vector $\boldsymbol{x}$.
Vector $\boldsymbol{c}$ and matrices $\boldsymbol{F}_{i}$ are the problem parameters.
The inequality $\boldsymbol{F}(\boldsymbol{x}) \succeq  0$ means that $\boldsymbol{F}(\boldsymbol{x})$ is a positive semi-definite matrix.

Any vector $\boldsymbol{x}$ satisfying the constraint $\boldsymbol{F}(\boldsymbol{x}) \succeq  0$ is called primal feasible point, and if $\boldsymbol{x}$ satisfies $\boldsymbol{F}(\boldsymbol{x}) \succ  0$, it is called strictly feasible point.
By convention, the minimal objective value $\boldsymbol{c}^{T} \cdot \hat{\boldsymbol{x}}$ is called primal optimal value.
Every primal problem has an associated dual problem.
Dual problem is a maximization problem and it is formulated as below
\begin{equation}
    \label{eq:SDPDual}
    \begin{aligned}
        \max\limits_{\boldsymbol{Z}} \quad &-Tr[\boldsymbol{F}_{0} \times \boldsymbol{Z}], \\
        s.t. \quad &\boldsymbol{Z}  \succeq  0, Tr[\boldsymbol{F}_{i} \times \boldsymbol{Z}]  =  \boldsymbol{c}_{i},
    \end{aligned}
\end{equation}
Here the optimization variable is the real symmetric (or Hermitian) positive matrix $\boldsymbol{Z}$ and the problem parameters are $\boldsymbol{Z}$ and $\boldsymbol{F}_{i}$ which are the same as in primal problem.
$Tr[\boldsymbol{A}]$ means trace of matrix $\boldsymbol{A}$.
Any matrix $\boldsymbol{Z}$ satisfying the constraints in dual problem is called dual feasible (or strictly dual feasible if $\boldsymbol{Z} \succ 0$).

The objective value of the primal feasible point is an upper bound on the minimal objective value $\boldsymbol{c}^{T} \cdot \hat{\boldsymbol{x}}$.
Similarly the objective value of the dual feasible point is an lower bound on the dual optimal value $-Tr[\boldsymbol{F}_{0} \times \hat{\boldsymbol{Z}}]$
For a primal feasible point $\boldsymbol{x}$ and a dual feasible point $\boldsymbol{Z}$ we have,
\begin{equation}
    \nonumber
    \begin{gathered}
        \boldsymbol{c}^{T} \cdot \boldsymbol{x}  +  Tr[\boldsymbol{F}_{0} \times \hat{\boldsymbol{Z}} ]  =  \sum_{i=1}^{|\boldsymbol{x}|} {  Tr[\boldsymbol{F}_{i} \times \boldsymbol{Z}] \boldsymbol{x}_{i}  }   +   Tr[\boldsymbol{F}_{0} \times \boldsymbol{Z}]  =  Tr[\boldsymbol{F}(x) \times \boldsymbol{Z}]  \geq  0.
    \end{gathered}
\end{equation}
This proves that $ -Tr[\boldsymbol{F}_{0} \times \hat{\boldsymbol{Z}}]   \leq   \boldsymbol{c}^{T} \cdot \hat{\boldsymbol{x}} $
and under relatively mild assumptions, we can have $ -Tr[\boldsymbol{F}_{0} \times \hat{\boldsymbol{Z}}]   =   \boldsymbol{c}^{T} \cdot \hat{\boldsymbol{x}} $.
If the equality holds, one can prove the following optimality condition on $\boldsymbol{x}$.

A primal feasible $\boldsymbol{x}$ and a dual feasible $\boldsymbol{Z}$ are optimal, if and only if
\begin{equation}
    \label{eq:SDPOptimality}
    \begin{gathered}
        \boldsymbol{F}(\hat{\boldsymbol{x}})  \times  \hat{\boldsymbol{Z}}  =  \hat{\boldsymbol{Z}}  \times  \boldsymbol{F}(\hat{\boldsymbol{x}})  =  0,
    \end{gathered}
\end{equation}
where $\hat{\boldsymbol{x}}$ and $\hat{\boldsymbol{Z}}$ are optimal ones.
This condition is called complementary slackness condition.

In one way or another, numerical methods for solving SDP problems always exploit the inequality
$-Tr[\boldsymbol{F}_{0} \times \boldsymbol{Z}]   \leq  -Tr[\boldsymbol{F}_{0} \times \hat{\boldsymbol{Z}}]   \leq  \boldsymbol{c}^{T} \cdot \hat{\boldsymbol{x}}  \leq  \boldsymbol{c}^{T} \cdot \boldsymbol{x}  $,
where $-Tr[\boldsymbol{F}_{0} \times \boldsymbol{Z}]$ and $\boldsymbol{c}^{T} \cdot \boldsymbol{x}$ are the objective values for any dual feasible point and primal feasible point, respectively.
The difference $\boldsymbol{c}^{T} \cdot \hat{\boldsymbol{x}}  +  Tr[\boldsymbol{F}_{0} \times \hat{\boldsymbol{Z}}]  =  Tr[ \boldsymbol{F}(\boldsymbol{x}) \times \boldsymbol{Z} ]  \geq  0  $ is called the duality gap.
If the equality  $    \boldsymbol{c}^{T} \cdot \hat{\boldsymbol{x}}  =  -  Tr[\boldsymbol{F}_{0} \times \hat{\boldsymbol{Z}} ]   $   holds, i.e. the optimal duality gap is zero, then it is said that strong duality holds.

\end{appendices}

\end{document}